\documentstyle[aps,prc,preprint]{revtex}
\tightenlines

\newcommand{\bea}{\begin{eqnarray}} 
\newcommand{\beq}{\begin{equation}} 
\newcommand{\ear}{\end{array}} 
\newcommand{\eea}{\end{eqnarray}} 
\newcommand{\eeq}{\end{equation}} 

\def\am{angular momentum\  }

\begin{document}

\draft

\title{\Large Description of rotating $N=Z$ nuclei in terms
of isovector pairing.}

\author{A.\ V.\ Afanasjev$^{(1,2)}$, S.\ Frauendorf$^{(1,3)}$} 

\address{$^{1}$Department of Physics, University of Notre Dame,
Notre Dame, Indiana 46556, USA}

\address{$^{2}$Laboratory of Radiation Physics, Institute of Solid State
Physics, University of Latvia, \\ LV 2169 Salaspils, Miera str. 31, Latvia}

\address{$^{3}$IKH, Research Center Rossendorf, Dresden, Germany}

\date{\today}

\maketitle

\begin{abstract}
A systematic investigation of the rotating $N=Z$ even-even nuclei in the
mass $A=58-80$ region has been performed within the frameworks of the 
Cranked Relativistic Mean field, Cranked Relativistic Hartree
Bogoliubov theories and cranked Nilsson-Strutinsky approach. Most 
of the experimental data is well accounted for in the calculations.
The present study suggests that there is strong isovector $np$-pair
field at low spin, the strength of which is defined by the isospin 
symmetry. At high spin, the isovector pair field is destroyed and 
the data are well described by the calculations assuming zero pairing. 
No clear evidence for the existence of the isoscalar $t=0$ $np$-pairing 
has been obtained in the present investigation. 

\end{abstract}


\input epsf

\section{Introduction}
\label{intro}

  It is well known that in the nuclei away from the $N=Z$ line  proton-proton 
($pp$) and neutron-neutron ($nn$) pairing dominate. In the $N\approx Z$ nuclei, 
protons and neutrons occupy the same levels. Strong $np$ pair correlations are 
expected because of large spatial overlap of their wave functions. These correlations 
can be isoscalar and isovector. Figuring out their character and whether 
they form a static pair condensate (an average field) in respective channel 
has been a challenge since medium mass  $N=Z$ nuclei have come into reach of 
experiment.

Shell model calculations show that there is strong isovector $np$-pairing the 
strength of which is fixed by isospin conservation \cite{engel,EPSVD.97}
and that all three isovector components ($nn,pp,np$) are rapidly suppressed by 
increasing the angular momentum \cite{DKLR.97}. The isoscalar correlations are 
weaker and do not change much with angular momentum. These results can be 
interpreted by the presence of a static  isovector pair field that is destroyed 
by rotation, like in nuclei far from the $N=Z$ line, and  by the existence of
dynamical isoscalar correlations that represent fluctuations around the zero 
mean value. The effective forces used in these calculations 
reproduce very well binding energies and other properties of $N\approx Z$ nuclei with $A<60$, 
and thus should well estimate the relative strength of the isoscalar and isovector
$np$ pair correlations. On the other hand, one may not exclude that these comparisons
are not sufficiently specific concerning $np$ pair correlations. The  analysis of pairing 
vibrations around $^{56}$Ni indicates  a  collective behavior of the isovector 
pairing vibrations but do not support any appreciable collectivity in 
the isoscalar channel \cite{bes-rev,MFCC.00}. Refs. \cite{alt,V.00,MFC.00} demonstrated that 
the relative energies of the lowest $T=0, 1/2, 1, 3/2, $ and $2$ states can be well accounted 
for by an isovector pair gap and a symmetry energy term proportional to $T(T+1)$. 
Having these results in mind, a natural approach seems starting with the assumption
that there is a strong isovector $np$ pair field, the strength of which is
determined by isospin conservation, and no isoscalar pair field. In this 
publication, we shall compare the rotational spectra of $N\approx Z$ 
nuclei with mean field calculations based on this assumption.

The strength of the isovector $np$-pairing is well defined by the isospin 
symmetry. A number of experimental observables such as binding energies of 
the $T=0$ and $T=1$ states in even-even and odd-odd $N=Z$ nuclei 
\cite{MFC.00,V.00,Rb74}, the observation of only one even-spin $T=0$ band 
in $^{74}$Rb \cite{Rb74} instead of two nearly degenerate bands expected 
in the case of no $t=1$ \footnote{The lower-case letter $t$ is used for the
isospin of the pair-field in order to avoid the confusion with the total 
isospin of the states denoted by $T$.} $np$-pairing clearly point on the 
existence of pairing condensate in this channel.

On the contrary, the strength of the isoscalar $t=0$ $np$-pairing is not
well known. Hence it is important to find physical properties that  are 
sensitive to it and may provide evidence for its presence. In a number of 
publications it was suggested that the  rotational properties of the 
$N\approx Z$ nuclei can serve for this purpose. However, in most of the 
cases these suggestions were based on the cranked  shell model (CSM) 
ignoring the considerable softness of the shape of the nuclei in the mass 
region of interest.
                                                                                
In the present manuscript, the cranked Nilsson-Strutinsky (CNS) approach 
\cite{Beng85,A110,PhysRep}, the cranked Relativistic Mean Field (CRMF) 
\cite{KR.89,KR.93,A150} and the cranked Relativistic Hartree-Bogoliubov 
(CRHB) \cite{A190,CRHB} theories (see Sect.\ \ref{theory}), which treat 
deformation properties more self-consistently than CSM, are employed 
for a detailed study of the rotating $N\approx Z$ nuclei 
in the mass range $A=64-80$. Together with previously published results 
\cite{Zn60SD,A60,Cu59}, they cover the mass range $A=58-80$. 
 These theories have succesfully been tested in a systematic way on the 
properties of different types of rotational bands such as normal-deformed 
\cite{J1Rare,A250} and superdeformed \cite{A60,A150,A190,CRHB} bands in 
the regimes of weak and strong pairing, as well as for smooth terminating 
bands \cite{A110,PhysRep,VRAL}. Thus, their accuracy for the nuclei away 
from the $N=Z$ line is well established with respect of which the accuracy 
of the description of the $N\approx Z$ nuclei can be judged. Compared with 
other theories, such as total routhian surface (TRS) 
calculations \cite{WS.01}, 
projected shell model (PSM) \cite{S.03} and complex EXCITED VAMPIR approach
\cite{PSF.02}, no adjustable parameters specific for nuclei in this mass 
region were used in our calculations.

  Although these theories do not take implicitly into account the $np$-pairing,
their use is justified by the study of the isovector mean field theory in Ref.\ 
\cite{FS.99-NP}. The isovector pair field breaks the isospin symmetry. Therefore, many 
solutions of the Hartree-Bogoliubov (HB) equations exist, which correspond to
different orientations of the pair field in isospace. One particular
orientation corresponds to the case of no $np$ pair-field. The $np$ pair 
correlations are taken into account by restoring the isospin symmetry by means of 
approximate methods that correspond
to the rotor or cranking models in the analogous case of breaking 
of  the angular momentum symmetry in deformed nuclei.

The article is organized as follows. In Section \ref{theory} the main features
of our theoretical tools are  outlined. Neccessary details of the calculations are
also given. In Section \ref{accuracy} the accuracy of the employed methods
is investigated for selected $N=Z+2$ nuclei.
The structure of even-even $N=Z$ with $64 \leq A \leq 80$ 
nuclei is investigated in Section \ref{spec-nucl}. 
The question whether there is evidence for the existence of an isoscalar 
$np$-pair field is discussed in Section \ref{disc}. Section \ref{concl} 
summarizes our main conclusions. 

\section{Theoretical tools}
\label{theory}
 
 The cranked Nilsson-Strutinsky approach \cite{Beng85,A110,PhysRep}, 
 the cranked Relativistic Mean Field \cite{KR.89,KR.93,A150} and the 
cranked Relativistic Hartree-Bogoliubov \cite{A190,CRHB} theories are employed 
in this manuscript for a detailed study of the rotating $N\approx Z$ nuclei. 
For high spin ($I\geq 15\hbar$) we neglect the pair correlations using the 
CRMF theory or the CNS approach. In the RMF approach the nucleus is described 
as a system 
of point-like nucleons represented by Dirac-spinors and coupled to mesons and 
to the photon. The nucleons interact by the exchange of several mesons, namely, 
the scalar $\sigma$ and three vector particles $\omega$, $\rho$ and the photon.
The CRMF theory represents the extension of RMF theory to the rotating frame.  
In the CNS approach the total energy is described as a sum of 
the rotating liquid drop energy and the shell correction energy. 
Although being much simpler than the self-consistent CRMF theory,
the CNS approach provides a reasonable description of the nuclear many-body 
problem \cite{PhysRep}. 

The CNS approach has several advantages compared to the 
self-consistent mean field models. These are the abilities (i) to specify 
a configuration in terms of
occupation of low- and high-$j$ orbitals, (ii) to trace fixed configuration 
up to the final termination in a non-collective  state, (iii) to 
study the same configuration at given spin in different local minima (such as 
collective and non-collective or positive and 
negative $\gamma$-deformation). Implementing them into the self-consistent theories 
requires the constraints on the configuration, spin and deformation over large 
deformation space.  As a result, such calculations are not numerically feasible
nowadays. Thus, in many cases the CRMF calculations (which are restricted to 
collective configurations only) are guided by the CNS results.

 CRMF and CRHB+LN calculations have been performed with the NL3 parameterization
of the RMF Lagrangian \cite{NL3} which provides rather good description of
nuclear properties throughout nuclear chart. The D1S Gogny force \cite{D1S} and
approximate particle number projection by means of the Lipkin-Nogami (LN) method
have been used in the pairing channel of the CRHB+LN theory. The CRMF 
and CRHB+LN equations are solved in the basis of an anisotropic three-dimensional 
harmonic oscillator in Cartesian coordinates with the deformation parameters 
$\beta_0=0.3$, 
$\gamma=0^{\circ}$ and oscillator frequency $\hbar \omega_0=41$A$^{-1/3}$ MeV.
All fermionic and bosonic states belonging to the shells up to $N_F=12$
and $N_B=16$ are taken into account in the diagonalization of the Dirac
equation and the matrix inversion of the Klein-Gordon equations, respectively.
The detailed investigation indicates that this truncation scheme provides 
good numerical accuracy. 

In order to investigate the dependence of the results on the parametrization
of the Nilsson potential, the CNS calculations have been performed with the
standard set of parameters \cite{Beng85} and the set suggested for the 
$A\sim 80$ mass region in Ref.\ \cite{GBI.86} ('A80' parameter set in the following).

 In the calculations without pairing, the shorthand notation 
$[p,n]$ indicating the number $p(n)$ of occupied $g_{9/2}$ proton 
(neutron) orbitals is used for labeling of the configurations. In the
cases when the holes in the $f_{7/2}$ subshell play a role, an
extended shorthand notation $[(p_h)p,(n_h)n]$ with $p_h(n_h)$
being the number of proton (neutron) $f_{7/2}$ holes is used. 
Since high-$j$ $f_{7/2}$ holes are important mainly in nuclei
around $^{60}$Zn \cite{A60}, in many cases we consider only mixed
low-$j$ $N=3$ orbitals and use $3_i$ label for them, where subscript
$i$ indicates the position of the orbital within the specific 
signature group.
An appreciable number of configurations should be considered when the 
calculations are performed as a function of $\omega$ in order to 
establish which configurations are lowest in energy at given spin 
and which ones have to be compared with experimental data.

\section{Accuracy of the mean-field description in the mass 60-80 region}
\label{accuracy}

It is well known that the shape of the nuclei in the mass 60-80 region 
changes strongly both with \am and with configuration. They are
characterized by considerable softness of potential energy surfaces 
\cite{NDBBR.85}. At high spin, superdeformation \cite{Zn60SD} and the 
termination of rotational bands \cite{Zn62bt,Br73} play an important role. 
These features must be taken into account when discussing the evidence for 
$np$ pairing. As examples for the dramatic shape changes and band termination
features encountered in this mass region we study $^{73,74}$Kr and $^{70}$Br
nuclei in the framework of the CNS and CRMF theories. Since it is expected 
that proton-neutron pairing is not important in the $N=Z+2$ nucleus $^{74}$Kr 
\cite{SatW.97}, the results for this nucleus provide a benchmark for the 
accuracy of the description of rotational properties within the CRHB+LN 
theory.

\subsection{ Unpaired regime: $^{74}$Kr.}
\label{74Kr-unpair}

\begin{figure}
\epsfxsize 16.0cm
\epsfbox{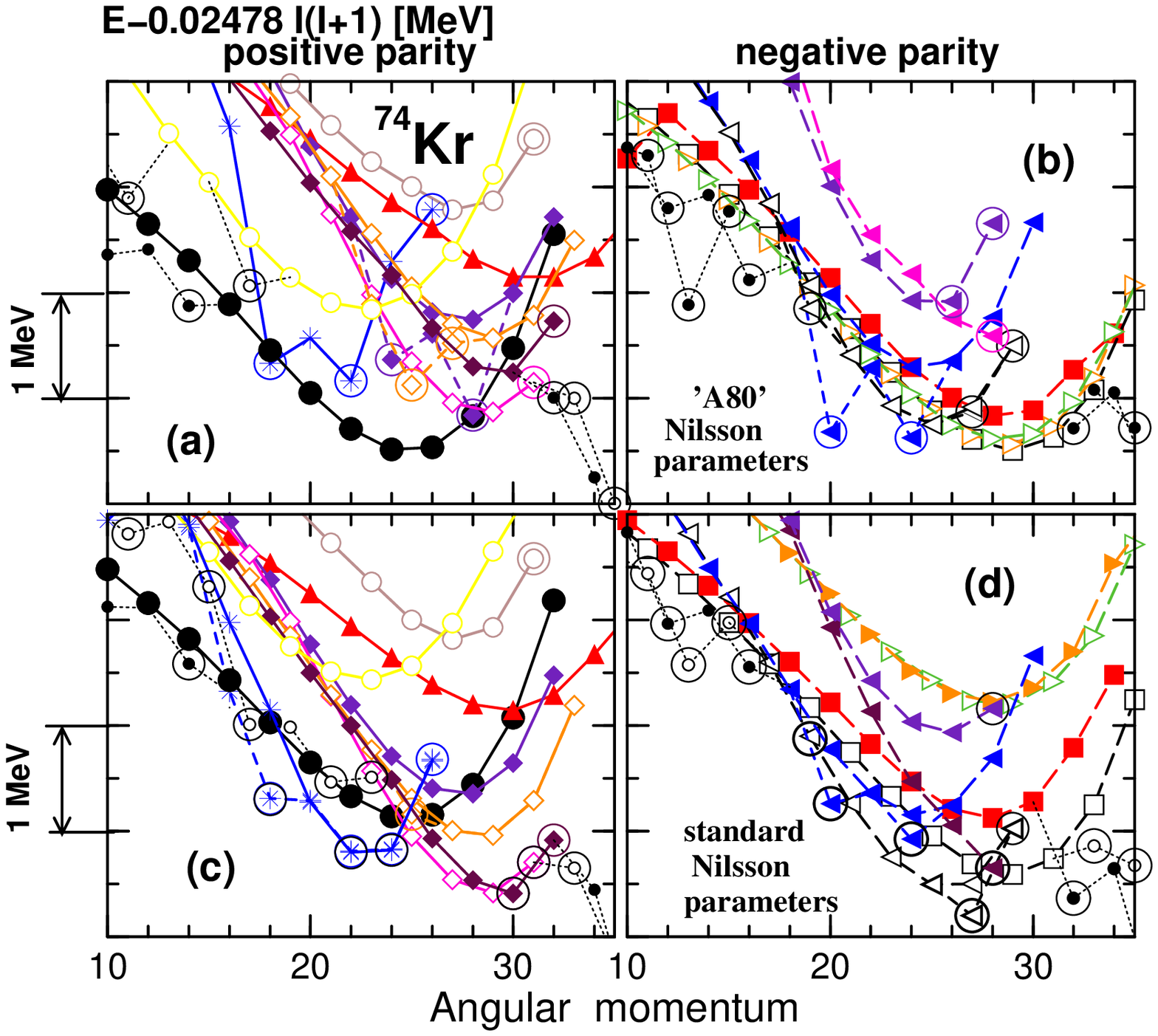}
\caption{Excitation energies of the $^{74}$Kr configurations relative to a 
rigid-rotor reference $E_{RLD}=0.02478 I(I+1)$ MeV obtained within the CNS 
approach. The standard set of the Nilsson parameters (bottom panels) is compared 
with the 'A80' set (top panels). Solid and open circles are used for 
signature $\alpha=0$ and $\alpha=1$ states, respectively. Solid and long-dashed 
lines are used for positive and negative parity states, respectively.  
Terminating (aligned) states are encircled. The absence of encircling indicates 
that the band continues beyond the maximum spin, see Refs.\ \protect\cite{PhysRep,A60} 
for details. Short-dashed lines are used to show non-collective and weakly 
collective branches of configurations. Dotted lines are used for 
yrast lines at low and high spins. 
The configurations correspond to the symbols: [2,2] - stars, [2,3] - triangles left,
[3,3] - diamonds, [2,4] - circles, [3,4] - squares, [4,4] -triangles up and
[(1)3,4] - triangles right.}
\label{kr74-a80+stand}
\end{figure}

Fig.\ \ref{kr74-a80+stand} shows the excitation energies of 
several configurations, forming the yrast line of four combinations 
of parity and signature, obtained in the CNS calculations with the
standard and 'A80' sets of the Nilsson parameters. Since the
two sets differ in the energies of various single-particle
orbitals, the relative energies of the  configurations 
depend strongly on the parametrization. For example, the proton 
$f_{7/2}$  spherical subshell is located too high in energy in the 
'A80' set. As a consequence, the signature degenerated [(1)3,4] configurations 
involving a hole in $\pi f_{7/2}$ compete with the [3,4] 
configurations (Fig.\ \ref{kr74-a80+stand}b). However, in
experiment such bands have not been observed (see discussion
below). The calculations with the standard Nilsson
parameters place such bands more than 1 MeV above the yrast line 
(Fig.\ \ref{kr74-a80+stand}d), in much closer correspondence with the 
 experimental situation. The [3,3] configurations, involving one neutron in the
$3_4(\alpha=\pm1/2)$ orbitals, are more energetically favored 
in the standard Nilsson set than in the 'A80' set, reflecting the different
energy spacing between the $\nu g_{9/2}$ orbitals and the above mentioned
orbitals in these sets. The experiment seems to favor the energy 
spacing between the orbitals obtained in the 'A80' set (see below).

  The CNS calculations (Fig.\ \ref{kr74-a80+stand}) predict the existence 
of a number of aligned states and the states with small collectivity 
($\gamma \sim 40^{\circ}$ and larger) along the yrast line. In addition, 
 collective and non-collective
minima coexist within the specific configurations 
 (see Sect.\ 6.5 in Ref.\ \cite{PhysRep} 
for a detailed discussion of such coexistence). The relative energies
between these minima depend strongly on the Nilsson parametrization
(for example, compare [2,2] configurations in Figs.\ \ref{kr74-a80+stand}a 
and c). At present, no aligned or weakly collective states of this
kind have been observed in $^{74}$Kr. Because of the
predicted small collectivity of these states as well as of their irregular
character, they are expected to be
much more difficult to be observed  than the more collective configurations. 
Although states of this kind competing with more 
regular collective bands have been predicted for a number of nuclei in the 
$A\sim 80$ mass region in different models \cite{NDBBR.85,PhysRep}, 
they have only been observed in $^{84}$Zr so far \cite{84Zr}.

\begin{figure}
\epsfxsize 16.0cm
\epsfbox{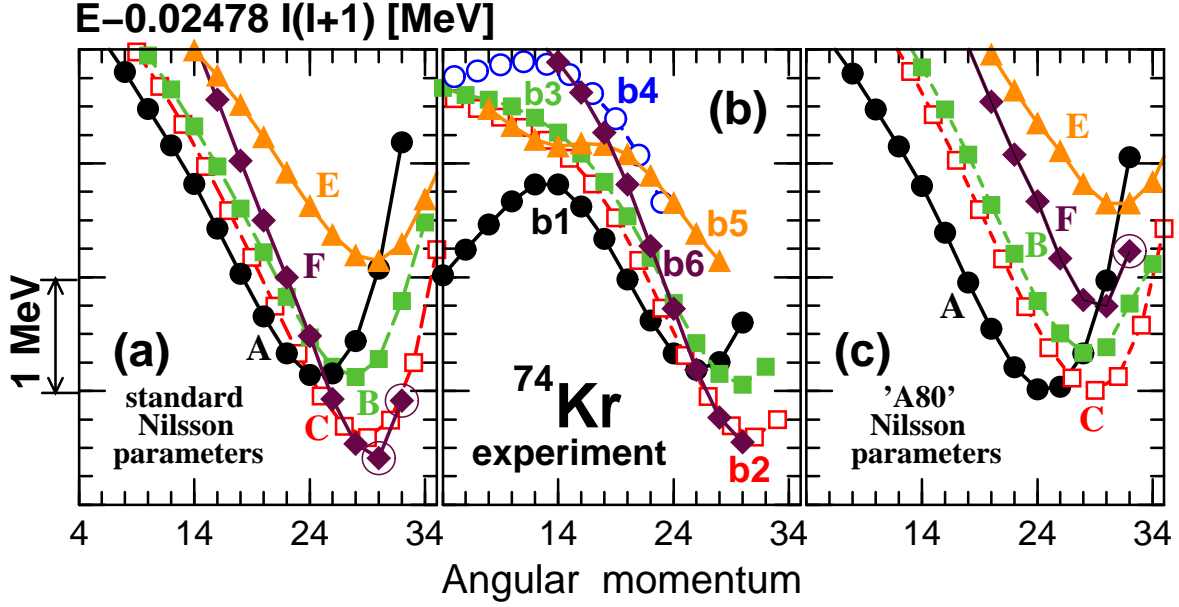}
\caption{Excitation energies of the experimental bands 1-6 in 
$^{74}$Kr  and the configurations [2,4] (A), [3,4] (B,C), [3,3] (F), 
and [4,4] (E) calculated in the CNS approach relative to a rigid  rotor 
reference $E_{RLD}=0.02478 I(I+1)$ MeV. Experimental data are taken from 
Refs.\ \protect\cite{Kr74,74Kr78Sr82Zr}. Note that the lowest state in 
band 6 is placed at an arbitrary excitation energy of $E_{exc}=4.9$ MeV and 
its spin is set to $I^{\pi}=8^+$. Solid and dashed lines are used 
for positive and negative parities, respectively. Solid and open symbols 
are used for signatures $\alpha=+1/2$ and $\alpha=-1/2$, respectively. 
In panel (b), solid circles, open squares, solid squares, open circles,
solid triangles, diamonds are used for bands 1, 2, 3, 4, 5 and 6, 
respectively.}
\label{kr74-eld-cns}
\end{figure}

\begin{figure}
\epsfxsize 16.0cm
\epsfbox{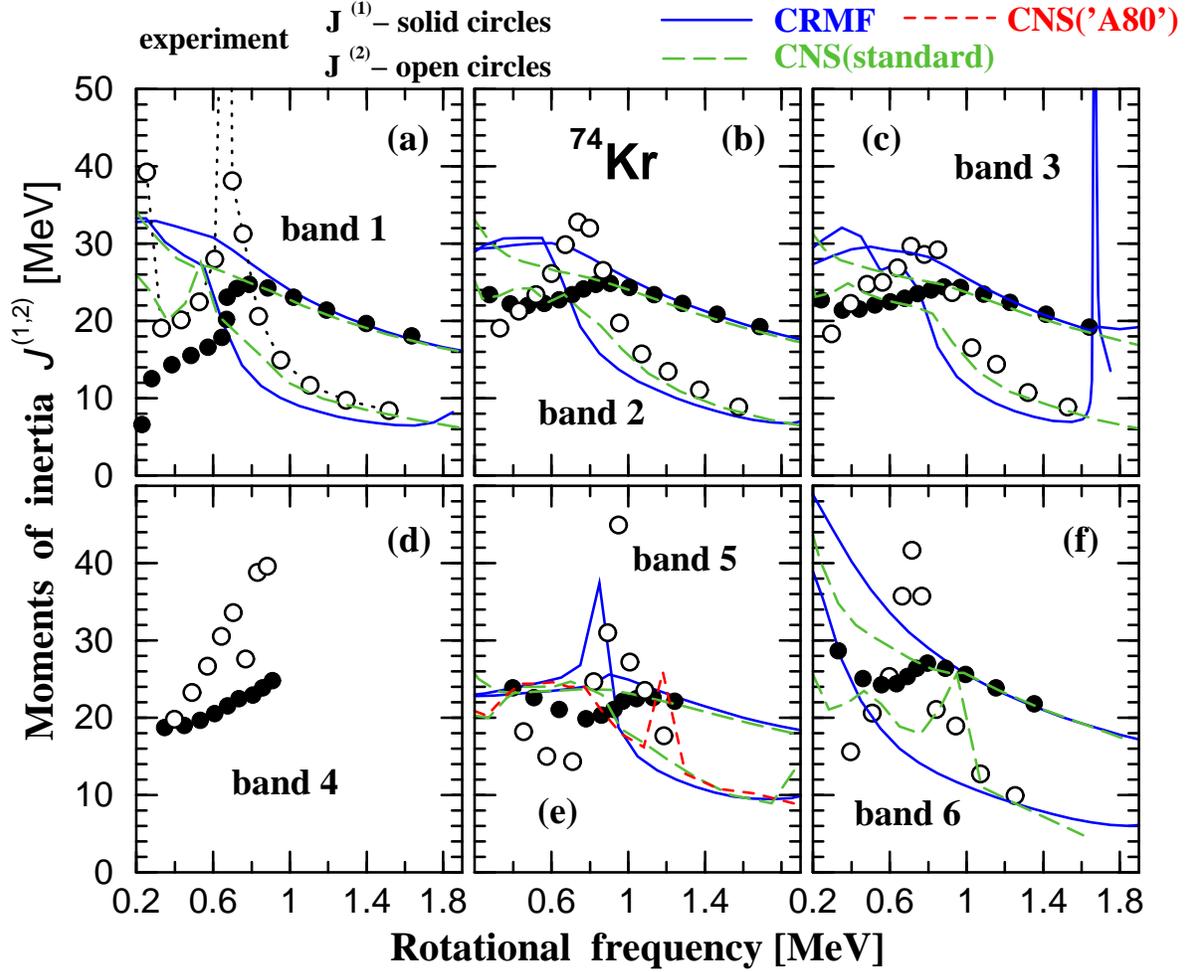}
\caption{Dynamic ($J^{(2)}$) and kinematic ($J^{(1)}$) moments of inertia 
of observed bands in $^{74}$Kr and corresponding calculated configurations. 
The results of the CNS and CRMF calculations are shown by different lines: 
the line being lowest at medium frequency corresponds to $J^{(2)}$, while 
highest one to $J^{(1)}$. The results of the CNS calculations with the 'A80' 
set of the Nilsson parameters are shown only when they differ considerably 
from those obtained with the standard parameters. Experimental data are 
taken from Refs.\ \protect\cite{Kr74,74Kr78Sr82Zr}.}
\label{kr74-j2j1}
\end{figure}

\begin{figure}
\vspace*{-1cm}
\epsfxsize 13.0cm
\epsfbox{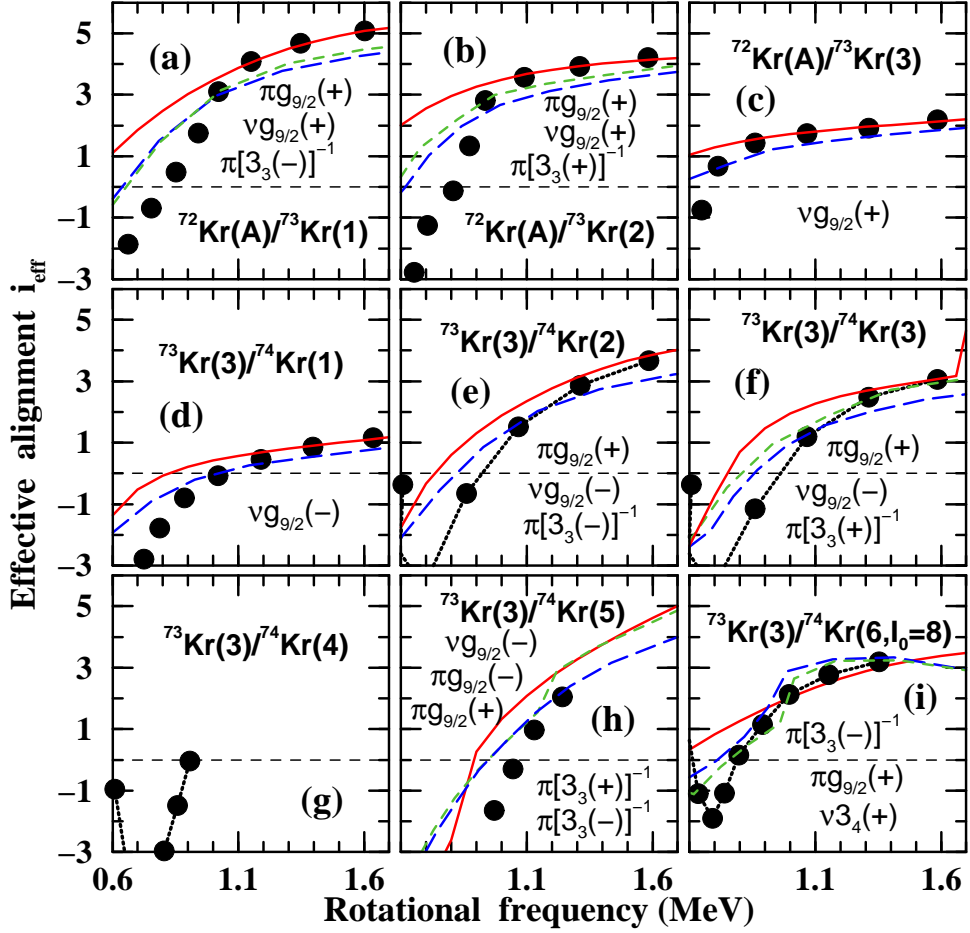}
\vspace*{1cm}
\caption{Effective alignments $i_{eff}$ (in units $\hbar$) of pairs of 
bands. The experimental values (solid circles) are compared with the
ones calculated from the configurations assigned
to the bands. The experimental effective alignment between bands
A and B is indicated as ``A/B''. The band A in the lighter
nucleus is taken as a reference so the effective alignment
measures the effect of additional particle(s)/hole(s). In order
to guide eye, the solid circles are linked by dotted lines
in a few cases. The
compared configurations differ by the occupation of the
orbitals indicated on the panels. The sign of signature 
$\alpha$ of the orbital is given in parentheses behind the
orbital label. $3_i$ refers to the $i$-th low-$j$ $N=3$
orbital, while superscript '$-1$' indicates the hole in
a given orbital. Solid, long-dashed and 
short-dashed lines are used for the effective alignments
obtained in the CRMF, CNS (standard Nilsson parameters)
and CNS ('A80' Nilsson parameters) calculations, 
respectively. The latter results are shown only if they
differ from the ones with the standard Nilsson parameters
by more than $0.25\hbar$. }
\label{kr-align}
\end{figure}

The rotational properties of specific collective configurations, reflected 
in the $(E-E_{RLD})$ curves (see Figs.\ \ref{kr74-a80+stand} and 
\ref{kr74-eld-cns}), the moments of inertia (Fig.\ \ref{kr74-j2j1}) and the 
effective alignments $i_{eff}$ (Fig.\ \ref{kr-align}), are less 
sensitive to the parametrization of the Nilsson potential. 
Thus, we use these properties for assigning 
 configurations to observed bands. These
assignments rely not only on the results of the CNS calculations, but 
also on the results of the CRMF calculations, which give a very similar
collective spectra. The minor differences are discussed below.

  The configuration $[2,4](\alpha=0)$ is assigned to positive parity band 1, 
while $[3,4](\alpha=0)$ and $[3,4](\alpha=1)$ to the negative parity bands 
3 and 2 (Fig.\ \ref{kr74-eld-cns}), respectively. The $[3,4](\alpha=0,1)$
configurations are created from $[2,4](\alpha=0)$ by an excitation of 
the proton from the $3_3(\alpha=\pm 1/2$) orbits into third 
$g_{9/2}$ orbit. The slopes, relative energies and the positions of the 
minima in the ($E-E_{RLD})$ plot (Fig.\ \ref{kr74-eld-cns}a,b), and, 
consequently, the dynamic and kinematic moments of inertia (Fig.\ 
\ref{kr74-j2j1}a-c) of bands 1-3 are very well reproduced by these 
configurations. The calculated effective alignments of band 3 in  
$^{73}$Kr(3) and bands 1, 2 and 3 in $^{74}$Kr  are close 
to experiment (Figs.\ \ref{kr-align}d,e,f) at rotational frequency 
above 1 MeV where  pairing is expected to be negligible.

These configurations are also the lowest collective configurations 
in the CRMF calculations. Concerning the 
excitation energies, the CRMF results are quite similar to the CNS 
ones shown in Fig.\ \ref{kr74-eld-cns}: the minima in the $(E-E_{RLD})$ 
curves are obtained at $I=26\hbar$, 
$I=30\hbar$ and $I=29\hbar$, respectively, in close agreement with experiment 
(see Fig.\ \ref{kr74-eld-cns}b). Also the relative energies, the slopes
of the $(E-E_{RLD})$ curves of bands 1-3 and the spins at which bands
2 and 3 cross band 1 are well reproduced by the CRMF calculations. 
At $\omega \geq 1.0$ MeV, the kinematic moments of inertia of these 
bands are well reproduced, while dynamic moments of inertia are 
somewhat underestimated (Fig.\ \ref{kr74-j2j1}a-c). The 
rise in $J^{(2)}$ in the configuration assigned to band 3
at $\omega \sim 1.65$ MeV (Fig.\ \ref{kr74-j2j1}c) is due to the 
crossing of the $\nu g_{9/2} (\alpha=-1/2)$ and 
$\nu (g_{7/2} d_{5/2})(\alpha=-1/2)$ orbitals.
The effective alignments of the band pairs $^{73}$Kr(3)-$^{74}$Kr(1,2,3)
are close to experiment (Figs.\ \ref{kr-align}d,e,f).

  Comparing experimental and calculated  $(E-E_{RLD})$ curves 
(Fig.\ \ref{kr74-eld-cns}) and effective alignments $i_{eff}$
(Fig.\ \ref{kr-align}h), we assign the unpaired
configuration $[4,4](\alpha=0)$ to  band 5 at high spin
(above $I=16\hbar$). While the  energy differences 
between band 5 and bands 2 and 3 are well reproduced in
both parametrizations, only the standard Nilsson parameters 
reproduce the excitation energy of band 5 with respect of 
band 1. Considering the differences in the configurations of
these bands (see Fig.\ \ref{kr-align}), one can conclude that
the energy gap between the $\pi 3_3(\alpha=\pm1/2)$ and 
$\pi g_{9/2}$ orbitals is much better described with the 
standard Nilsson parameters than with the 'A80' ones. This is 
clearly seen in the relative energies of bands 1, 2 and 3 in 
$^{74}$Kr (Fig.\ \ref{kr74-eld-cns}) and bands 1, 2 and 3 in 
$^{73}$Kr (see Fig.\ 6 in Ref.\ \cite{73Kr} for the results with 
the 'A80' set), which are sensitive to the energy gap between 
the above mentioned orbitals. These relative energies are better 
reproduced by the standard Nilsson parameters.

   Our configuration assignment of  band 5 agrees with that
obtained within the TRS framework \cite{74Kr78Sr82Zr}, where it was 
suggested that the lowest 2qp positive-parity band (band 5) is built 
on the $(\pi g_{9/2})^2$ (AB) configuration which undergoes band 
crossing at $\omega \sim 0.9$ MeV due to the alignment of the 
$(\nu g_{9/2})^2$ neutron pair. 

  The interpretation of band 6 is more ambiguous, because it 
is not linked to the low-spin level scheme. Using the effective 
alignment, we suggest that it is based on the $[3,3]
(\alpha=0)$ configuration (see Fig.\ \ref{kr-align}i) and its lowest 
state has spin $I_0=8\hbar$. Assuming this spin, the slope of experimental 
$(E-E_{RLD})$ curve is well reproduced (Fig.\ \ref{kr74-eld-cns}). The 
weak point of this interpretation is the fact that the $[3,3](\alpha=1)$ 
configuration is predicted to have a  lower  
energy (especially in the 'A80' set) than this 
configuration (Fig.\ \ref{kr74-a80+stand}a and c). However, both in the 
CNS calculations with the standard parameters and in the CRMF calculations 
it is lower in energy by only few hundred keV in the  short spin 
range $I=23-27\hbar$. This 
difference in predicting the relative 
energies of the lowest $[3,3](\alpha=0,1)$
configurations can be traced to the signature splitting of the 
$\nu 3_4(\alpha=\pm 1/2)$ orbitals. 

 The fact that the $[3,3](\alpha=\pm 1/2)$ configurations have been observed 
in $^{73}$Kr (bands 1 and 2) \cite{73Kr}, strongly suggests the presence of 
similar configurations in $^{74}$Kr. Indeed, by adding 
one $3_4 (\alpha=\pm 1/2)$  
neutron to these configurations of $^{74}$Kr, four [3,3] configurations
 are created
(Fig.\ \ref{kr74-a80+stand}a and c). In the CRMF calculations, these [3,3]
configurations are somewhat less energetically favored with respect of 
the [2,4] configuration used as a reference in comparison with the CNS calculations.

  Since band 4  does not reach the 
region of weak pairing (Fig.\ \ref{kr74-j2j1}d), an unambiguous interpretation
of this band in the formalism without pairing is difficult. However, since
the  signature 
partner of this band has not been observed, the signature degenerated [(1)3,4] 
configurations can be excluded. Thus, only decoupled $[2,3](\alpha=1)$ 
configuration (Fig.\ \ref{kr74-a80+stand}b and d) seems to be a reasonable 
candidate for the extension of this band to high spin. The analysis of the CRMF 
results leads to the same conclusion.

\subsection{Unpaired regime: $^{73}$Kr}
\label{Kr73-sect}

  This nucleus has been studied in detail in Ref.\ \cite{73Kr} by means 
of the CRMF theory and the CNS approach employing the 'A80' Nilsson parameters. 
In addition, we have carried out a CNS study  with the standard parameters for 
the Nilsson potential,  which gave similar results, where, however, the relative 
energy of band 3 and  bands 1 and 2 is better reproduced (see Sect.\ 
\ref{74Kr-unpair}). At $\omega \geq 1.0$ MeV, the calculations without pairing 
(CNS and CRMF) reproduce well experimental data; see Figs.\ 6 and 7 in Ref.\ 
\cite{73Kr}.

\subsection{Softness of the $N\sim Z$ nuclei: $^{70}$Br}
\label{Br70-soft}

\begin{figure}
\epsfxsize 16.0cm
\epsfbox{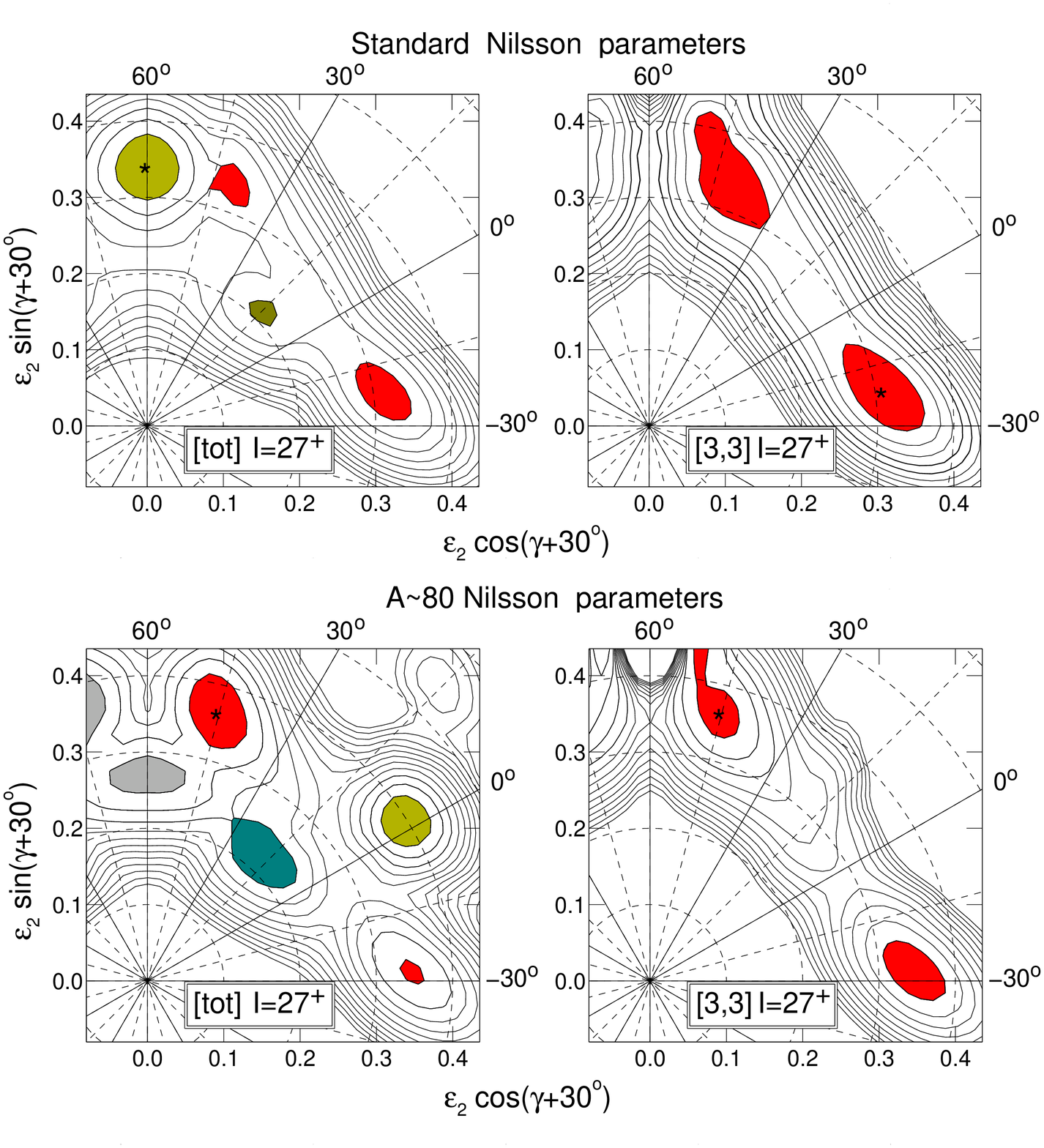}
\caption{Potential energy surfaces in $^{70}$Br for  spin $I=27^+$. 
In the left panels  only parity ($\pi=+$) and signature ($\alpha=1$) 
are fixed. In the right panels only the [3,3] configuration is shown. 
The contour line separation is 0.25 MeV. The last equipotential 
line corresponds to 3.0 MeV excitation with respect to the global 
minimum. The local minima, the excitation of which with respect of the 
global minimum does not exceed 1.0 MeV, are shaded. The results with 
the standard and 'A80' Nilsson parameters are presented in top and 
bottom panels, respectively.}
\label{comp-pes-br70-a80-stand}
\end{figure}

 Selected potential energy surfaces of $^{70}$Br  in Fig.\ 
\ref{comp-pes-br70-a80-stand} illustrate the soft nature of the $N=Z$ 
nuclei in the $A\sim 70$ mass region. They also show that the height 
of the barrier between the minima at $(\varepsilon_2\approx 0.35, 
\gamma \sim -20^{\circ})$ and $(\varepsilon_2\approx 0.35, \gamma 
\sim 30^{\circ})$ depends on the parametrization of the Nilsson 
potential. On the other hand, the relative energies of these two 
minima and their deformations are less sensitive to the parametrization 
of the Nilsson potential. 

\begin{figure}
\epsfxsize 16.0cm
\epsfbox{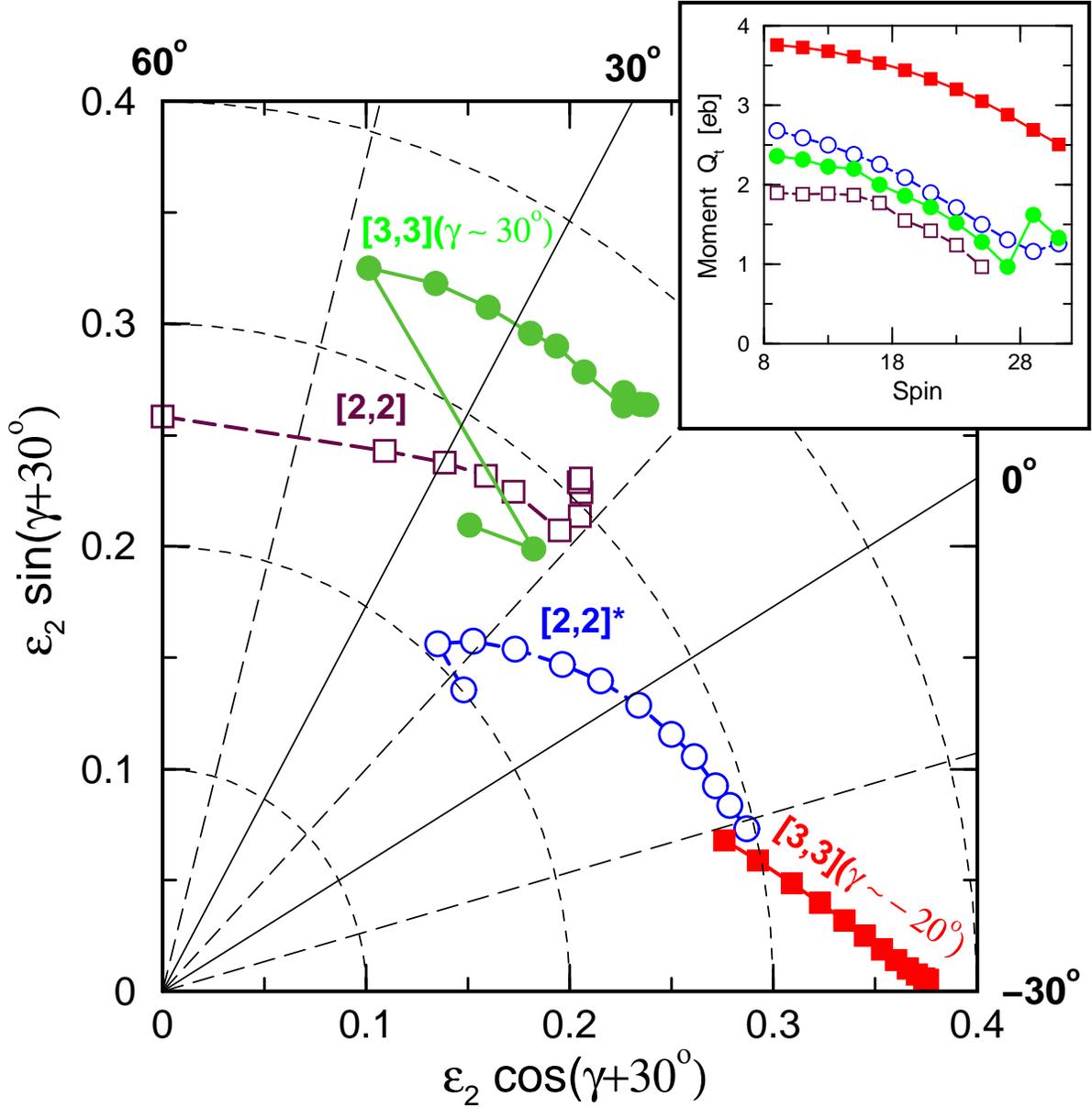}
\caption{The calculated shape trajectories of the [2,2] and [3,3]
configurations in $^{70}$Br. The corresponding transition quadrupole 
moments $Q_t$ are shown in the insert. The shape trajectories are shown
for $9\hbar \leq I \leq  31\hbar$ in steps
of 2$\hbar$.
The [2,2] configuration terminates at $I=27\hbar$. The initial
point on right side of each trajectory corresponds to the lowest spin.
}
\label{br70-defpath}
\end{figure}

In the $N=Z$ nuclei with particle numbers 34, 35, and 36 the observed 
bands are associated with 
either the [2,2] or [3,3] configurations, residing in these two competing 
local minima, or both of them. The high spin part of the band HB1 in $^{70}$Br 
is the envelope of the [3,3] configurations \cite{Br70}.
The CNS calculations also indicate the existence of  collective [2,2] 
configurations associated with similar two local minima (see Ref.\ \cite{Br70} 
for details and Fig.\ \ref{br70-defpath}), but they have not been observed 
in experiment. The [2,2] configurations dominate the yrast line in $^{68}$Se 
because of the lower Fermi level. The bands belonging to
 such a structure 
have been observed (see Sect.\ \ref{Se68-sect}). The bands 
with [2,2] and [3,3] configurations have been observed in $^{72}$Kr 
(see Sect.\ \ref{Kr72-sect}).

\subsection{ CRHB+LN theory: $^{74}$Kr}
\label{74Kr-crhb+ln}

\begin{figure}
\epsfxsize 16.0cm
\epsfbox{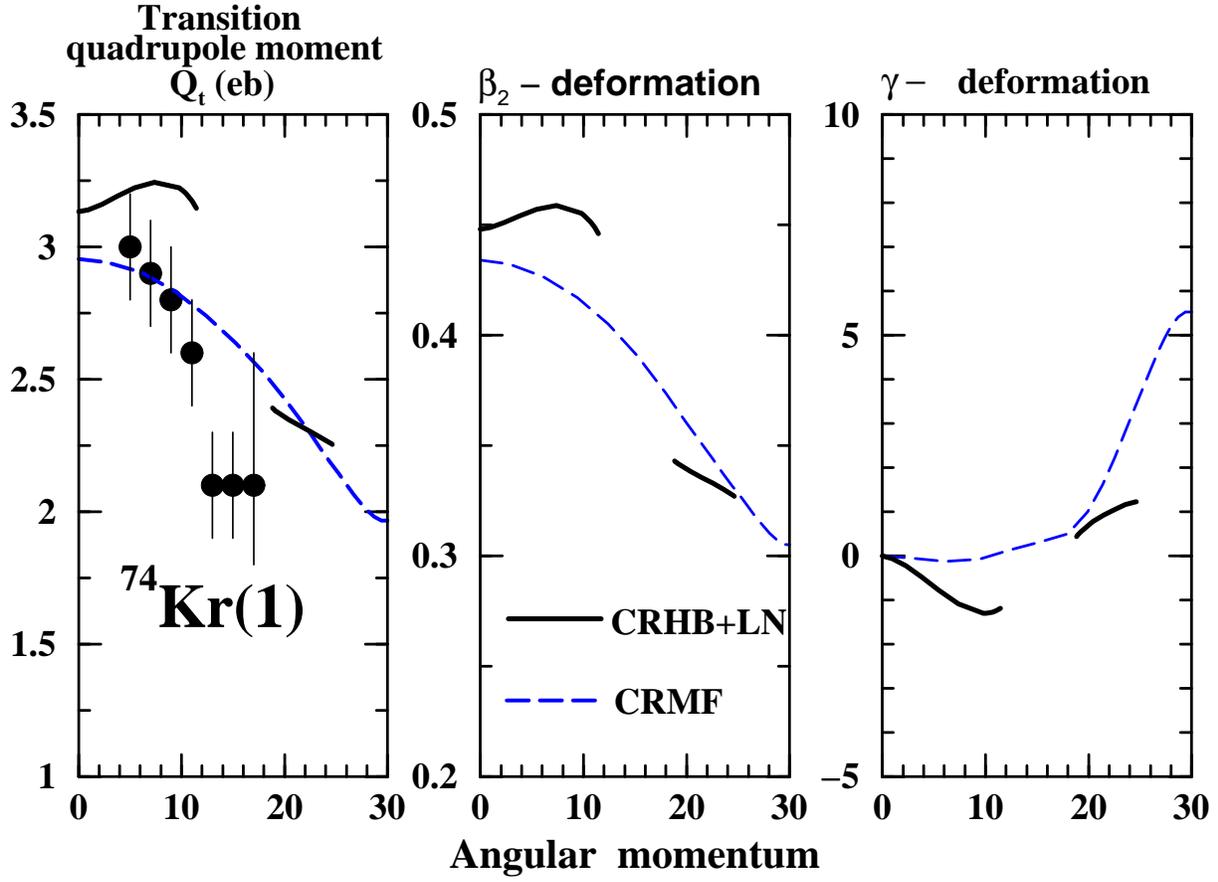}
\caption{ Deformation properties of the ground state band in $^{74}$Kr. 
Experimental data are taken from Ref.\ \protect\cite{74Kr-Qt}. The results 
of the CRHB+LN and CRMF (for the [2,4] configuration) calculations are 
shown by solid and dashed lines, respectively.}
\label{kr74-def}
\end{figure}

Fig.\ \ref{kr-pair-j1}b compares the kinematic moment of inertia of 
$^{74}$Kr calculated within the CRHB+LN theory with the experimental data. 
With exception of the lowest frequency point and two points after band 
crossing the experimental data is excellently reproduced by the configuration 
that is lowest in energy in the near-prolate minimum. The lowest frequency 
point is well reproduced by the oblate configuration,  which suggests
a transition from oblate to prolate shape at $I \sim 4\hbar$ within 
the ground state band of $^{74}$Kr. 
The band crossing is 
caused by the simultaneous alignment of the proton and neutron $g_{9/2}$ pairs.
No convergence has been obtained for $\omega \geq 1.21$ MeV due to very weak 
pairing at these frequencies. However, the similarity of the CRHB+LN and CRMF 
results for $J^{(1)}$ above the band crossing suggests that the weak $t=1$ 
pairing only insignificantly modifies the rotational properties. The same 
result has been obtained  in $^{72}$Kr (Fig.\ \ref{kr-pair-j1}a) 
and in $^{60}$Zn (Ref.\ \cite{Pingst-A30-60}). The CRHB+LN calculations 
reproduce well the transition quadrupole moment $Q_t$ of the ground state band 
and its dramatic drop after the band crossing (see Fig.\ \ref{kr74-def}). These 
calculations together with the CRMF calculations for the [2,4] configuration 
(unpaired analog of the S-band) indicate a gradual decrease 
of $Q_t$ with increasing spin. It is caused by a decrease of $\beta_2$
and an increase of $\gamma$-deformation.

\subsection{$^{78}$Sr}
\label{78Sr-sect}

   A delay of the first band crossing in the ground state band of even-even
$N=Z$ nucleus with respect to the one in the $N=Z+2$ nucleus has been widely
discussed as an evidence for the $t=0$ $np$-pairing \cite{Fisher}. However,
in order to apply this argumentation the band crossing frequency has to be well 
established in the $N=Z+2$ system. While this is unproblematic in $^{74}$Kr 
(see Sect.\ \ref{74Kr-crhb+ln}), the situation in $^{78}$Sr is more complicated. 
The detailed discussion below illustrates this. 

To facilitate the discussion of positive parity bands of $^{78}$Sr (see Fig.\ 3 
in Ref.\ \cite{74Kr78Sr82Zr}) we use the label A for the branch consisting of the 
ground state band up to $I=12^+$ state, and the states at the energies 6025 
($14^+$), 7559 ($16^+$), 9254 ($18^+$), 11195 ($20^+$) and 13294 ($22^+$) keV. The 
label B is used for the branch consisting of the states at the energies 9254 ($18^+$), 
10995 ($20^+$), 12981 ($22^+$), 15233 ($24^+$) and 17764 ($26^+$) keV.

\begin{figure}
\epsfxsize 14.0cm
\epsfbox{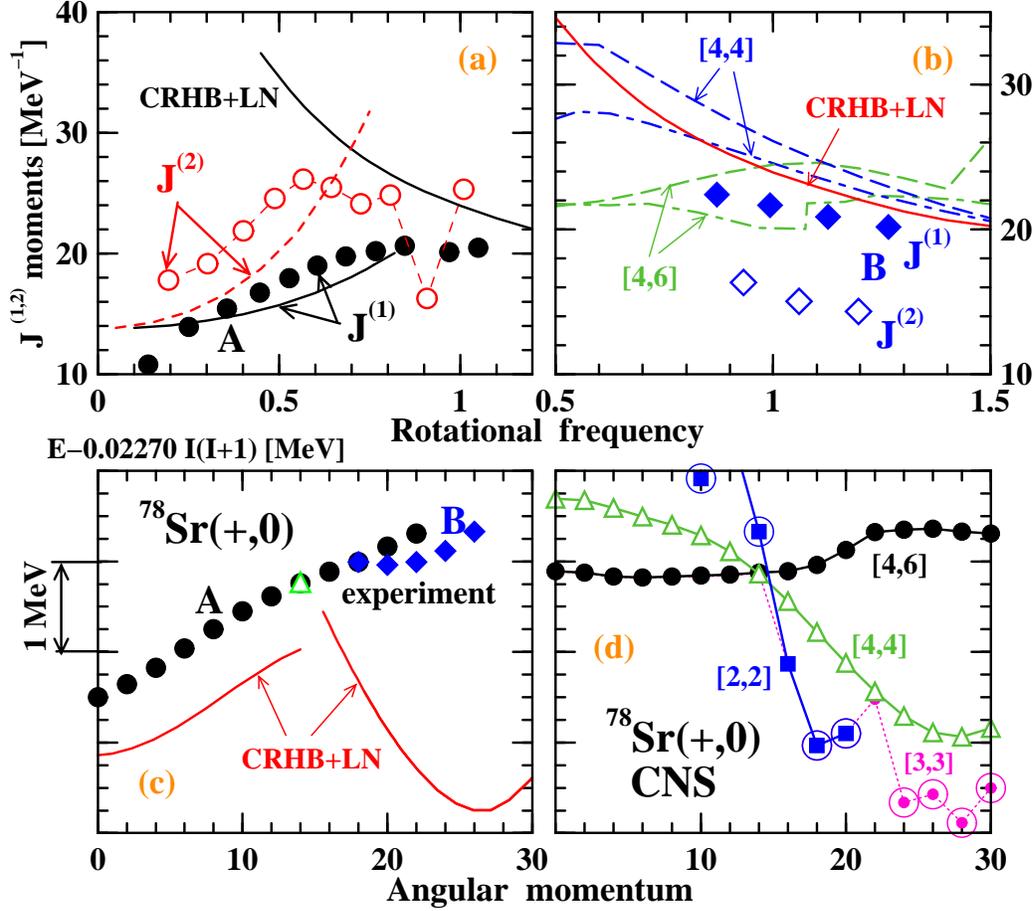}
\caption{Panel (a): Kinematic (solid symbols) and dynamic (open linked 
symbols) moments of inertia of branch A in $^{78}$Sr 
compared with the CRHB+LN results.
Calculated $J^{(1)}$ and $J^{(2)}$ (only up 
to back-bending) values are shown by solid and dashed lines, respectively.
Panel (b): the same as in panel (a) but for the branch B. 
The CRHB+LN, CRMF and CNS results for $J^{(1)}$ are shown by solid, 
long-dashed and dot-dashed lines, respectively. (c) Excitation energies 
of the $(\pi=+, \alpha=0)$ structures (unlinked symbols) compared 
with the results of the CRHB+LN calculations (solid lines). They are given 
with respect of the rigid rotor reference. (d) The CNS results for the 
yrast and near-yrast collective and non-collective structures
with $(\pi=+, \alpha=0)$ shown in the same way as in 
panel (c). The yrast line is indicated by a dotted line.}
\label{sr78}
\end{figure}

  The results of the CRHB+LN and CNS calculations and experimental data are shown 
in Fig.\ \ref{sr78}. The branch B shows all features typical for the rotation in 
the unpaired regime such as $J^{(2)}\leq J^{(1)}$ and the smooth decrease of both 
quantities with increasing rotational frequency (Fig.\ \ref{sr78}b). This 
smoothness strongly suggests that the branch B is not affected by the interaction 
with another band. Note that the state at 9254 ($18^+$) keV is included in both 
branches. The smoothness of $J^{(2)}$ and $J^{(1)}$ with this state included 
in branch B (see Fig.\ \ref{sr78}b) suggests that this state belongs to band B 
not to band A. It is likely that the branch B is the continuation of the 
g-band above the point of crossing with the S-band at $I\sim 16\hbar$ (see below) and 
corresponds to the unpaired [4,6] configuration (compare panels (c) and (d) of Fig.\ 
\ref{sr78}). However, neither CRMF not CNS calculations describe well the kinematic 
moment of inertia of branch B (Fig.\ \ref{sr78}b) under this configuration assignment. 
Similar problems with the description of this branch has been encountered also in the 
TRS calculations of Ref.\ \cite{74Kr78Sr82Zr}.

  The dip in $J^{(2)}$ of branch A at $\omega \sim 0.9$ MeV (Fig.\ \ref{sr78}a), which 
is not typical for paired band crossing, suggests that at this frequency the g-band 
is crossed by some other configuration. This means that the states at 11195 keV ($20^+$) 
and at 13294 keV ($22^+$) of the branch A do not  belong  to the g-band. We 
were not able to make configuration assignment for these two states.

  The presence of two closely lying $14^+$ states does not allow a certain assignment 
of one of them to the g-band although the intensities within the bands suggest that
the state at 6025 MeV belongs to g-band. Thus we can only be more or less certain that  
the states up to $16^+$ in branch A belong to the g-band. In this range, the CRHB+LN 
calculations reproduce well the experimental $(E-E_{RLD})$ plot (Fig.\ \ref{sr78}c) 
and the moments of inertia (Fig.\ \ref{sr78}a; see also discussion below). 

  Extrapolating the calculated CRHB+LN $(E-E_{RLD})$ curve of the g-band to the crossing 
point with the S-band predicts a back-bend at $I\sim 16\hbar$ (Fig.\ \ref{sr78}a). The 
calculated equilibrium deformation of S-band ($Q_t \sim 2.37$ $e$b, $\beta_2 \sim 0.28$ 
and $\gamma \sim -20^{\circ}$) differs considerably from the one of the g-band 
($Q_t \sim 3.6 $ $e$b, $\beta_2 \sim 0.47$ and $\gamma \sim -1^{\circ}$). After simultaneous 
alignment of the proton and neutron $g_{9/2}$ pairs, the pairing is weak in the S-band, 
and thus this band can be accociated with the unpaired [4,4] configuration. A similar 
situation is found also in the CNS calculations where the [4,6] configuration (unpaired 
analog of the g-band) is crossed by the [4,4] configuration (unpaired analog of the S-band) 
at $I=14\hbar$ (Fig.\ \ref{sr78}d). The deformations of these configurations are 
similar to those obtained in the CRHB+LN calculations for the g- and S-bands. In addition, 
the CNS calculations indicate that the yrast line above $I=14\hbar$ is formed by the weakly 
collective or non-collective aligned states. A configuration similar to the S-band of 
the CRHB+LN calculations has also been obtained in the TRS calculations of 
Ref.\ \cite{74Kr78Sr82Zr}.

   The CNS calculations suggest that one of two closely lying $I=14^+$ states
is either the aligned state of the [2,2] configuration (Fig.\ \ref{sr78}d) 
or the state of the [4,4] configuration (S-band) (Fig.\ \ref{sr78}c and d).
The interaction between these two states might explain the discrepancy 
between the calculated and experimental $J^{(2)}$ values for g-band at 
$\omega \sim 0.7$ MeV (Fig.\ \ref{sr78}a).

   The present results strongly suggest that the S-band has not been observed 
in $^{78}$Sr, and, thus the use of this nucleus as a reference point in comparing 
band crossing frequencies is not justified. The situation is reminiscent to $^{72}$Kr
in the past, where the  S-band was missing in the early experiments (see Sect.\ 
\ref{Kr72-sect}). Additional experimental studies at $I\geq 14\hbar$ are needed 
in order to better understand the structure of $^{78}$Sr.

   The conclusion about non-observation of the S-band in $^{78}$Sr is strongly 
supported by the analysis of both the complicated level scheme in neighboring 
$^{77}$Rb nucleus, which has one proton less and the g- and S-bands in $^{76}$Kr 
which has two proton less (see Fig.\ \ref{kr-pair-j1}). In $^{77}$Rb, the high 
spin bands P3 \cite{77Rb-1,77Rb-2} and N2 \cite{77Rb-2} are associated with the 
[3,4] and [3,5] configurations, respectively, and their properties are well 
described by the CNS calculations \cite{AR-unp}. For example, the crossing of 
two ($\pi=+, \alpha=+1/2$) bands P3 and P1 
seen at $I=18\hbar$ in experiment is well described by the crossing of 
the [3,6] and [3,4] bands. However, pairing is still playing a role 
in the band P1 so its low spin properties are not completely reproduced 
in the calculations without pairing. It is clear that one additional proton 
in the $g_{9/2}$ orbital, leading to the [4,6] and [4,4] configurations 
in $^{78}$Sr discussed above, should not modify considerably the situation 
as have been seen in the Kr isotope chain (see Sects.\ \ref{74Kr-unpair}, 
\ref{Kr73-sect}, \ref{Kr72-sect}).

    Our analysis reveals the problems in interpretating
this $N=Z+2$ nucleus by means of our theoretical methods. At present, it is not 
clear to which extend they are due to the deficiencies of the models 
or to incomplete data. However, this study clearly indicates that
one has to be very careful to relate  possible
problems in understanding the experimental data on  $N=Z$ nuclei
to the lack of  $t=0 ~np$-pairing in the  
models.

\subsection{Band termination}
\label{bt-feat}

\begin{figure}
\epsfxsize 16.0cm
\epsfbox{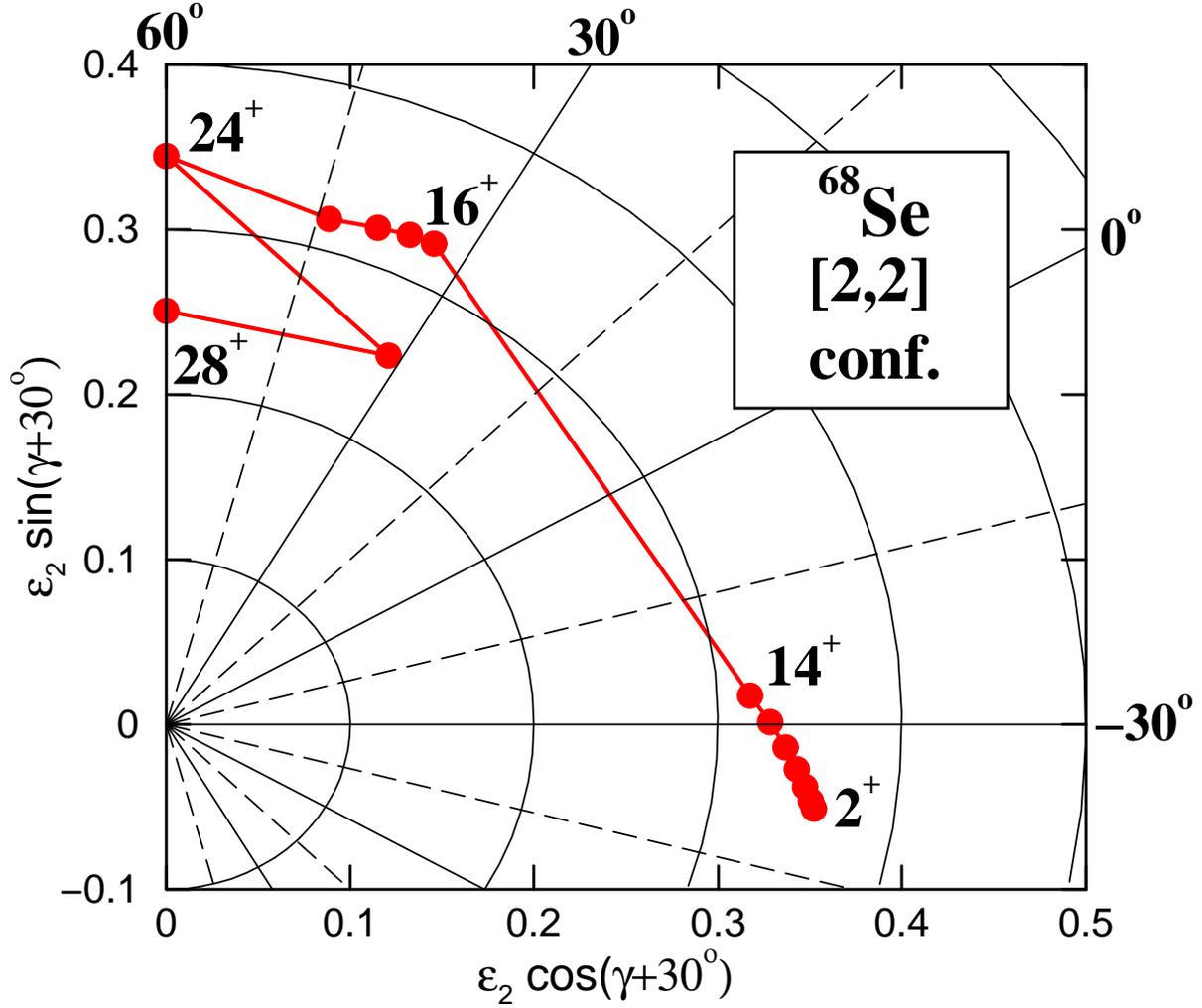}
\caption{The calculated (CNS) shape trajectory of the [2,2] 
configuration in $^{68}$Se. The angular momentum changes  in steps
 of 2$\hbar$ in the indicated spin range.}
\label{se68-defpath}
\end{figure}

    Above the rotational frequency of paired band crossings at $\omega = 0.6-1.0$ 
MeV, almost all experimental bands studied in this paper show the same feature: a 
drop of the dynamic moment of inertia below the kinematic one and decreasing kinematic 
moment of inertia with increasing rotational frequency. Such a behaviour is typical 
for smooth band termination \cite{PhysRep}, which is caused by limited angular momentum 
content of underlying single-particle configurations. Band termination corresponds to 
a gradual change from collective (near-prolate, triaxial) shape towards non-collective 
oblate shape, as illustrated in Fig.\ \ref{se68-defpath}. As documented in Ref.\ 
\cite{PhysRep}, the CNS approach describes reliably such bands in nuclei 
away from the $N=Z$ line. As shown in the present manuscript the CNS and CRMF approaches 
describe the high-spin bands in the $N\approx Z$ nuclei with a comparable level of 
accuracy.

\subsection{Single-particle spectra}
\label{sp-accur}

The uncertainties in the energies of the single-particle states will affect
the shell effects, which determine
the positions of local minima in deformation plane and the 
barriers between these minima (see Sect.\ \ref{Br70-soft}). The rotational 
properties and relative energies of configurations and their mutual 
interaction will also be affected. These properties will be less sensitive to 
inaccuracies of the single particle spectrum 
if the gap between  interacting states is large. For example, 
the effective alignments of bands in $^{72,73,74}$Kr which differ by the 
occupation of the $\pi 3_3(\alpha= \pm 1/2)$ orbitals are more sensitive to the 
parametrization of 
the Nilsson potential than those between the bands differing in the occupation 
of the $\pi g_{9/2}$ subshell (Fig.\ \ref{kr-align}), because  
the relative energies of $\pi 3_2(\alpha = \pm 1/2)$ and $\pi 3_3(\alpha= \pm 1/2)$ orbitals and their 
interaction depend strongly on the parametrization.

\subsection{Concluding remarks}
 
 The study of selected $N=Z+2$ nuclei presented in this Section reveals 
a number of features typical for $N\approx Z$ nuclei with 
$60<A\leq 80$. It illustrates the typical accuracy of the  theoretical 
tools employed. The potential energy surfaces of these nuclei are 
shallow  and shape coexistence is a common feature. The rotational bands 
show all the features of terminating bands at high spin. The process of band 
termination is associated with a drastic shape change from near-prolate 
(triaxial) shape at low spin towards non-collective oblate shape at high 
spin. The transition from the g-band to S-band is associated with a  
shape change as well. These phenomena cannot be ignored 
in studying the consequences of $np$-pairing for the properties of rotating 
nuclei. Our analysis also demonstrates  that the results of the calculations 
depend on how accurately the energies of the single-particle states are reproduced.
Hence, certain  discrepancies between 
calculations and experiment seen in the $N=Z$ nuclei (Sect.\ \ref{spec-nucl}) 
should not necessarily be attributed to the neglect of $t=0$ $np-$pairing.

\section{The structure of even-even $N = Z$ nuclei}
\label{spec-nucl}

The application of the isovector mean-field theory \cite{FS.99-NP}
to even-even $N=Z$ nuclei is very simple. Since all low-lying rotational bands 
have isospin $T=0$, the calculations by means of the CRHB+LN model can be directly 
applied to rotational bands in these nuclei, because the isorotational 
energy term $T(T+1)/2{\cal J}_{iso}$ vanishes. On the level of accuracy of the standard 
mean-field calculations, the restoration of the isospin symmetry (which takes care of 
the t=1 $np$ pair field) changes only the energy of the $T=1$ states relative to the $T=0$ 
states \cite{FS.99-NP}. The high spin states are systematically studied by means 
of the CSN and CRMF approaches, which assume zero pairing. Calculation of the low and
intermediate spin states by means of the CRHB+LN theory for selected nuclei 
complement the study.

\subsection{$^{64}$Ge }
\label{Ge64-sect}

It has been pointed out before that this nucleus is soft with respect to
$\gamma-$ and octupole deformations (see Ref.\ \cite{Ge64-exp} and references 
quoted therein). Fig.\ \ref{Ge64-eld} shows the results of the CNS calculations
(restricted to reflection symmetric shapes), which also indicate  softness  
toward  $\gamma$-deformation. The 
[0,0]$(\alpha=0)$ configuration is characterized by the $(\varepsilon_2 \sim 0.2,
\gamma \sim  -30^{\circ})$ deformation in the spin range $I=2-8\hbar$. The yrast 
lines of other combinations of parity and signature are characterized by similar 
deformation in the spin range $I=0-3\hbar$. Up to spins $I\sim 35\hbar$, the
yrast lines are dominated by the states with the deformations $\varepsilon_2 
\approx 0.25-0.35$ and $\gamma=26^{\circ}-60^{\circ}$. Band terminations,
mostly in a favored way \cite{PhysRep}, are typical for the yrast
region up to $I\sim 35\hbar$. Superdeformed bands with deformation 
$\varepsilon_2 \sim 0.5, \gamma \sim 10^{\circ}$ become yrast above that spin.

It seems to us that the complicated structure of this nucleus, dominated in the 
spin region of interest by $\gamma-$ (and probably octupole) softness and 
terminating structures, will not allow to obtain reliable evidences of the isoscalar 
$t=0$ $np-$pairing even if the experimental data will be extended to higher 
spin.

\subsection{$^{68}$Se nucleus}
\label{Se68-sect}

\begin{figure}
\epsfxsize 12.0cm
\epsfbox{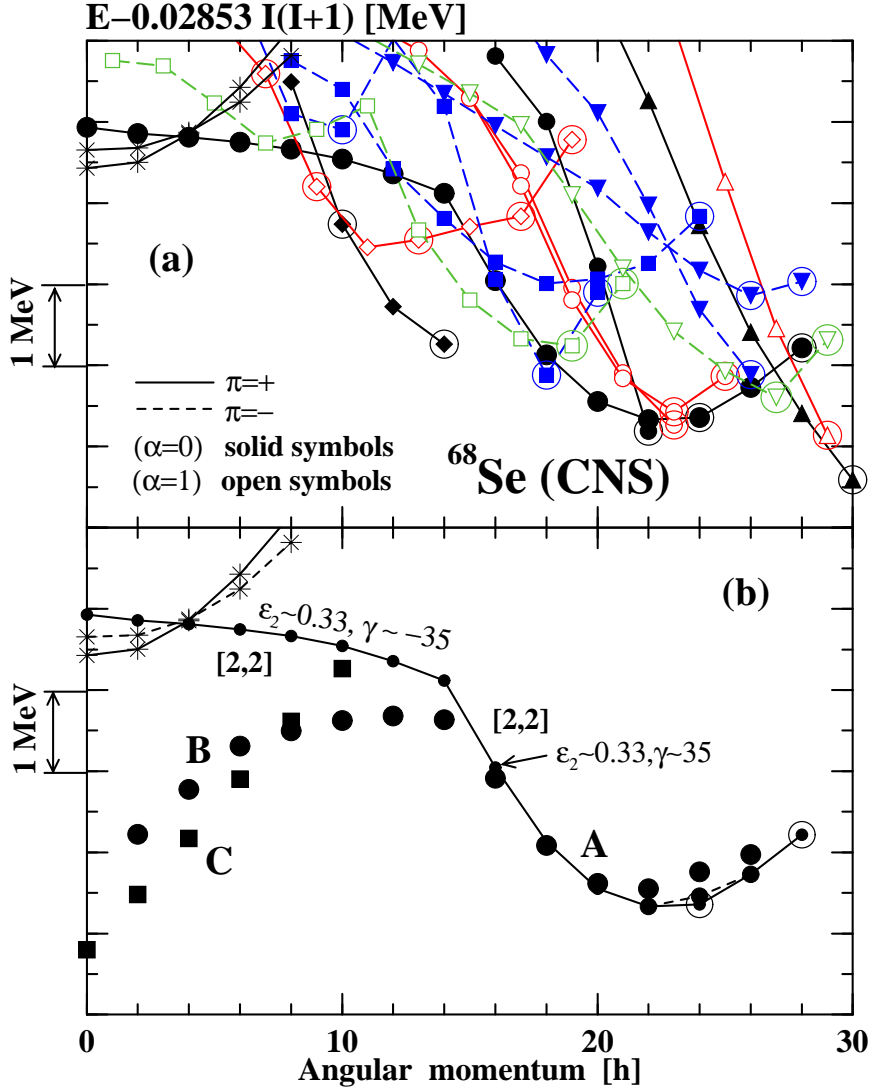}
\caption {(a) Excitation energies of the configurations forming the 
yrast line relative 
to a rigid rotor reference $E_{RLD}=0.02853 I(I+1)$ MeV. The calculated 
terminating (aligned) states are encircled. Only the yrast 
$(\pi=+,\alpha=0)$ configurations are shown for $I\leq 10\hbar$.
Different symbols are used for different types of configurations:
stars - [0,0], diamonds - [1,1], squares - [2,1], circles - [2,2],
triangles down - [3,2], triangles up -[21,2]. (b) The same as panel 
(a) but for experimental bands shown by unlinked symbols and assigned 
configurations shown by lines with symbols. Squares and circles 
are used for rotational sequences C and A/B, respectively. The
data are taken from Ref.\ \protect\cite{68Se72Kr}.}
\label{se68-eld}
\end{figure}

  As shown in Fig.\ \ref{se68-eld}a, the structure of $^{68}$Se calculated
by means of the CNS approach is quite complicated with different configurations 
competing for yrast status. However, the positive parity $[2,2](\alpha=0)$ 
configuration (shown by solid line with large solid circles) stays close 
to yrast line over a considerable spin range. Also the CRMF calculations
indicate that this configuration is energetically favoured in the spin range 
of interest. However, it is yrast in the $(\pi=+,\alpha=0)$ group of states 
only within some spin range. For example, the configurations $[1,1](\alpha=0)$ 
and $[21,2](\alpha=0)$ compete with it for yrast status at spins $I=10-14\hbar$ 
and $I=28\hbar$, respectively. 

  Fig.\ \ref{se68-eld}b compares the $[2,2](\alpha=0)$ configuration
with the rotational sequences A and B  observed in this nucleus
\cite{68Se72Kr}. For $I > 14\hbar$, the calculations are
in almost perfect agreement with band A. The state at $I=16\hbar$ has a 
deformation of $\epsilon_2=0.326$ and $\gamma=33^{\circ}$. With increasing 
spin the quadrupole deformation generally decreases but $\gamma$-deformation 
increases towards $\gamma=60^{\circ}$ (Fig.\ \ref{se68-defpath}). This is 
due to the 
limited angular momentum content of the 
configuration which leads to band termination  \cite{A110,PhysRep}. 
The CNS calculations give two aligned states  within 
this configuration. The state at 
$I=28\hbar$, which corresponds to maximum 
spin, has  the structure
$\pi (g_{9/2})^2_8 (f_{5/2}\,p_{3/2})^4_6 \otimes \nu (g_{9/2})^2_8 (f_{5/2}\,p_{3/2})^4_6$.
The state at $I=24\hbar$ has the structure 
$\pi (g_{9/2})^2_8 (f_{5/2}\,p_{3/2})^4_4 \otimes \nu (g_{9/2})^2_8 (f_{5/2}\,p_{3/2})^4_4$
with a different alignment of the low-$j$ $N=3$
protons and neutrons (see Sect. 6.5 of Ref.\ \cite{PhysRep} for a discussion
of a similar situation in $^{158}$Er). At $I=24\hbar$, the potential energy 
surface indicates the coexistence of the non-collective aligned and 
more collective highly triaxial states within the $[2,2](\alpha=0)$ 
configuration.

\begin{figure}
\epsfxsize 17.0cm
\epsfbox{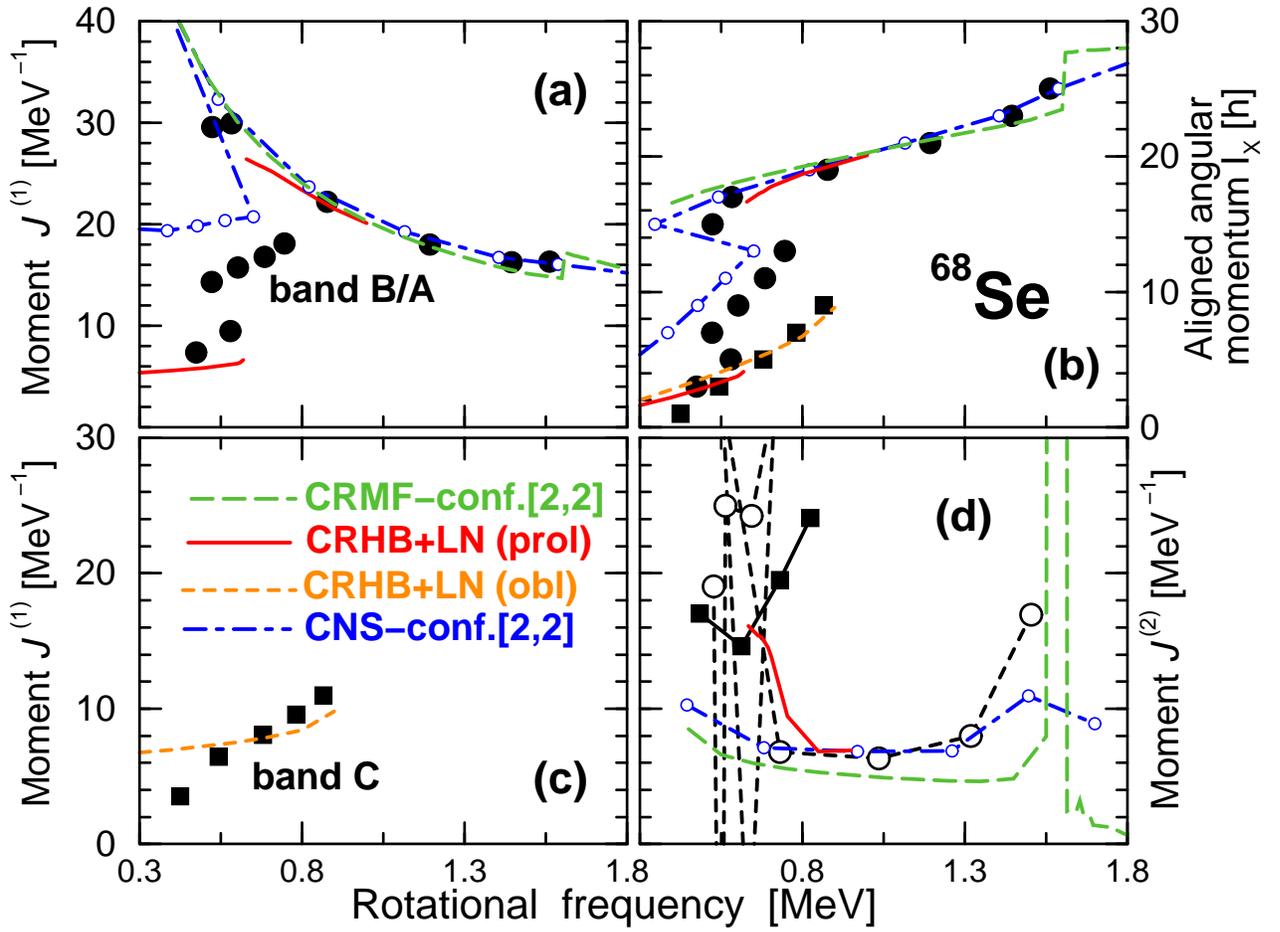}
\caption{Kinematic moments of inertia (panels a,c),
aligned angular momentum (panel b), and dynamic moments of inertia (panel d) 
in $^{68}$Se as  functions of 
rotational frequency. The data  from Ref.\ \protect\cite{68Se72Kr} 
are compared with the results of CRHB+LN (in prolate and oblate minima),
CRMF (only in the minimum with positive $\gamma$) and CNS calculations.
The CRHB+LN results in the oblate minimum (panel (c)) are shown only up 
to the frequency of the band crossing ($\omega =0.91$ MeV).}
\label{se68-j2j1}
\end{figure}

 For $\omega \approx 0.7 - 1.3$ MeV,  the dynamic 
moment of inertia of band A is essentially flat (Fig.\ \ref{se68-j2j1}d) and it 
is significantly lower than the kinematic moment of inertia which
decreases with increasing rotational frequency (Fig.\ \ref{se68-j2j1}a). 
This is a feature typical for rotational bands in unpaired regime 
approaching the limit of angular momentum that can be generated by the
valence particles and holes (such as smooth terminating bands
\cite{A110,PhysRep} and highly- and superdeformed bands in the
$A\sim 60$ mass region \cite{A60}). At $\omega \sim 1.4$ MeV, 
there is a rise in $J^{(2)}$. The CRMF calculations  show 
that this is due to the crossing of two low-$j$ $r=+i$ $N=3$ 
single-particle orbitals both in proton and neutron subsystems. 
The CNS approach gives the same explanation for the rise of 
$J^{(2)}$.

 The calculations indicate the coexistence of collective
highly triaxial and noncollective aligned states at spin 
$I=24^+$ within the $[2,2](\alpha=0)$ configuration. The analysis 
suggests that the observed $I=24^+$ state in band A is the triaxial 
collective state shown by dashed line in Fig.\ \ref{se68-eld}b. 
It is reasonable to expect that the intensity of the $26^+ \rightarrow 24^+$ 
and $24^+ \rightarrow 22^+$ transitions will be much weaker if the 
$24^+$ state is aligned as compared with the case when this state is 
collective. This is one possible reason why aligned $I=24^+$ state
has not been observed. It may also be that in reality this aligned 
state is less favoured in energy than predicted by the CNS. 
The terminating state at  $I=28^+$ has not been observed either.
In the CNS calculations this state lies 0.8 MeV above the yrast 
line which suggests that it may not be well populated in experiment. 
The CRMF calculations also indicate the termination of the $[2,2](\alpha=0)$ 
configuration at $I=28\hbar$ in the non-collective
$\gamma=60^{\circ}$ state, which is seen in Fig.\ \ref{se68-j2j1}b 
as the fact that the increase of $\omega$ above $1.6$ MeV does not  
change $I$.

  The CNS calculations suggest that the top part of the band B ($ 8<I<14\hbar$)
 corresponds to the collective 
$(\epsilon_2 \sim 0.33, \gamma \sim -35^{\circ})$ $[2,2](\alpha=0)$ 
configuration (Fig.\ \ref{se68-eld}b). The slope of the experimental
$(E-E_{RLD})$ curve in this spin range is reasonably well reproduced,
although the excitation energy is somewhat overestimated. The reason
for the last discrepancy could be twofold: neglect of pairing correlations
and the accuracy of the description of the energy difference
between the minima with ($\varepsilon_2 \sim 0.33, \gamma \sim 33^{\circ}$)
and $(\varepsilon_2 \sim 0.33, \gamma \sim -35^{\circ})$.
The inclusion of pairing will decrease the excitation
energy of the ($\varepsilon_2 \sim 0.33, \gamma \sim -35^{\circ}$)
branch of the $[2,2](\alpha=0)$ configuration and will
bend the  $(E-E_{RLD})$ curve toward the experimental one. 
In addition, it will decrease the kinematic moment of 
inertia and aligned angular momentum towards the values seen in experiment 
(Fig.\ \ref{se68-j2j1}a and b). Despite the neglect of pairing the 
transition from negative-$\gamma$ collective branch to the terminating 
branch at $I\sim 14\hbar$ (Fig.\ \ref{se68-eld}b) as well as the 
general features of kinematic moment of inertia (Fig.\ \ref{se68-j2j1}a) 
and aligned angular momentum (Fig.\ \ref{se68-j2j1}b) 
 are well reproduced in the CNS calculations in the frequency range above 
0.5 MeV.

 In general, band B for $I>8\hbar$ and band
A have a  structure very similar to the high-spin band (HB1) 
in $^{70}$Br. In Ref.\  \cite{Br70}, this band was interpreted as built
from two $[3,3]$ configurations. The low- and medium- spin part of
band was assigned to the $(\varepsilon_2 \sim 0.35, 
\gamma \sim -20^{\circ})$ local minumum, while top part of band to 
the $(\varepsilon_2 \sim 0.35, \gamma \sim 30^{\circ})$ minimum. Thus band 
B/A in $^{68}$Se and HB1 band in $^{70}$Br differ merely by 
one $g_{9/2}$ proton and one $g_{9/2}$ neutron, but they are 
located in the same local minima of potential energy surface.
Similar [2,2] configurations  as in $^{68}$Se have
also been predicted for $^{70}$Br but have not been observed so
far (see Ref.\ \cite{Br70} for details).

 In order to understand the low-spin structure, CRHB+LN 
calculations have been performed. At $I=0\hbar$, the oblate minimum 
is lower than the prolate one by 0.8 MeV in very close agreement 
with experiment \cite{Se68-old}. If the pairing is 
neglected  the energy difference between two minima is
smaller, because  the  pairing energy is larger in 
oblate minimum (1.67 MeV)  than in prolate one (1.28 MeV).
The CRMF calculations give a highly-deformed triaxial [2,2] 
configuration with $(\beta_2 \sim 0.42, \gamma \sim -20 ^{\circ})$
which corresponds to the one in the CNS approach. At $I=0\hbar$, 
it has an excitation energy of 3.2 MeV with respect of the 
lowest state in the oblate minimum. At $I\sim 14\hbar$, it is  crossed 
by the terminating branch of the [2,2] configuration. A  comparison of the 
CRMF and CRHB+LN results for the prolate and oblate minima indicates that 
pairing has small effect on the equilibrium deformation. Assuming that 
this is also true for the highly-deformed triaxial minimum and that 
the gain in binding due to pairing is similar to the one  in prolate
and oblate minima, the  above mentioned excitation energy is not far from 
the energy of band B extrapolated to spin $I=0\hbar$, which is somewhat 
larger than 2 MeV.

 The kinematic moments of inertia and aligned angular momentum of band C and the bottom 
branch of band B (spins $I=2-6\hbar$) are compared with the CRHB+LN results 
in Fig.\ \ref{se68-j2j1}a,b and c. The calculations describe the experiment 
reasonably well, although some discrepancies are seen at $\omega \leq 0.55$ 
MeV. Due to convergency problems caused by weak pairing, the CRHB+LN 
calculations do not extend beyond 1.0 MeV. It is clearly seen that the 
CRHB+LN and CRMF results converge for $\omega \geq 0.8$ MeV 
(Fig.\ \ref{se68-j2j1}a). Both the prolate CRHB+LN solution 
($\beta_2 \sim 0.24, \gamma \sim -3^{\circ}$)
and the oblate one ($\beta_2 \sim 0.26, \gamma 
\sim -60^{\circ}$) are crossed  by a 
band that contains an aligned  $g_{9/2}$ proton pair and an aligned $g_{9/2}$  neutron pair
($\beta_2 \sim 0.34, \gamma 
\sim 25^{\circ}$), which corresponds to the 
unpaired $[2,2](\alpha=0)$ configuration.

The structure of this nucleus has  been studied before. Apart from minor details, the 
TRS interpretation of bands C, B and of the bottom of band A \cite{WS.01} 
is in agreement with the one given above. However, the termination of band A 
has not been investigated. Projected shell model calculations for the  oblate and prolate 
minima were performed in Ref.\ \cite{SMAW.03}. The interpretation of band C and bottom 
part of band B (spins $I=2-6\hbar$) is the same as in our calculations. However, the 
middle part of band B (spins $I=8-14\hbar$) is interpreted 
differently as  either proton or neutron $K=1$ 
two-quasiparticle bands (in unpaired language they would correspond to 
either [2,0] or [0,2]). According to our CNS and CRMF 
calculations, these configurations are more excited 
(by more than 1 MeV in CNS and by few MeV in CRMF) than the $[2,2]$ 
configuration with negative $\gamma$-deformation and thus should not be 
assigned  to the $I=8-14\hbar$ branch of band B. 
CNS and CRMF calculations also show  that  in the spin range 
$I=10-14\hbar$ the configuration $[1,1]$ is somewhat lower in energy 
than [2,2]. However, we deem this very unlikely.

The complex version of the Excited Vampir variational approach
which includes neutron-proton pairing has been applied for the
study of the even-spin positive parity $^{68}$Se spectra up to 
$I=16^+$ in Ref.\ \cite{PSF.02}. A good description of the energies 
has been obtained by  adjusting several 
parameters. In particular, the isoscalar spin 0 and 1 particle-particle 
matrix elements were enhanced and all   
diagonal  T=0 monopole matrix elements  were shifted. This shift turns out to be  
important for reproducing the prolate-oblate energy difference, which is obtained in the 
CRHB+LN calculations without any adjustable parameters. These calculations
interpret band C and bottom part of band
B similar to us, but the 
interpretation of top part of band B (spins $I=8-14\hbar$)
is different. It remains to be seen whether
configurations corresponding to the highly-deformed negative-$\gamma$ minimum
have to be included in the Excited Vampir calculations 
and in projected shell model for a correct interpretation of this part 
of band B.

 Our interpretation could be corroborated by measurements of transition quadrupole 
moments $Q_t$. According to the CRHB+LN calculations, band C and the low-spin 
part of band B ($I=2-6\hbar$) have average values of $Q_t\approx $1.75 $e$b and 1.65  
$e$b, respectively, which are slightly increasing with spin. 
The $I=8-14\hbar$ part of band B is characterized by quite 
large values of $Q_t \approx 2.65$ $e$b [CRMF] and $Q_t \approx 3.05$ $e$b 
[CNS], if it is associated with negative-$\gamma$ $[2,2](\alpha=0)$ 
configuration. If this branch, however, has the [1,1] structure, then 
the CRMF calculations give $Q_t\approx 1.4$ $e$b. For band  A, CRMF predicts 
that the $Q_t$ values decrease from 1.4 $e$b at $I=16\hbar$  to 0.95 $e$b 
at $I=26\hbar$. On the other hand, CNS predicts a decrease from  1.3 $e$b
at $I=16\hbar$ to  0.8 $e$b at $I=22\hbar$.

\subsection{$^{72}$Kr}
\label{Kr72-sect}

 The alleged delay of the crossing between the g-band and  
the doubly aligned S-band in $^{72}$Kr has been in the focus of considerations 
relating it to the $np$-pairing. 
Thereby it was assumed that the high spin part of band B (in the notation 
of Ref.\ \cite{68Se72Kr}) represents the doubly aligned S-band (see 
Ref.\ \cite{Fisher} and references therein). The rational was that the 
mean-field calculations predicted this crossing 
at about the same frequency as in the adjacent isotopes with N=74 and
76, whereas band B shows only a gentle upbend at a substantially 
higher frequency.  However the situation changed with the 
observation of a second even spin band (band A) of positive parity in Refs.\ 
\cite{Berk-2,72Kr-epj}, which was then confirmed in Ref.\ \cite{68Se72Kr}. 
The only difference between these studies is the  
energy of the  $(26^+)\rightarrow (24^+)$ transition. In Refs.\ 
\cite{Berk-2,72Kr-epj} it has an energy of (2837) keV, while Ref.\ 
\cite{68Se72Kr} suggests an energy of (3063) keV and assignes the 2834 keV line 
(which is very close to 2837 keV transition of 
Ref.\ \cite{Berk-2,72Kr-epj}) to the link between bands A and B. In the present 
manuscript, we follow the level scheme of Ref.\ \cite{68Se72Kr}.

\begin{figure}
\epsfxsize 16.0cm
\epsfbox{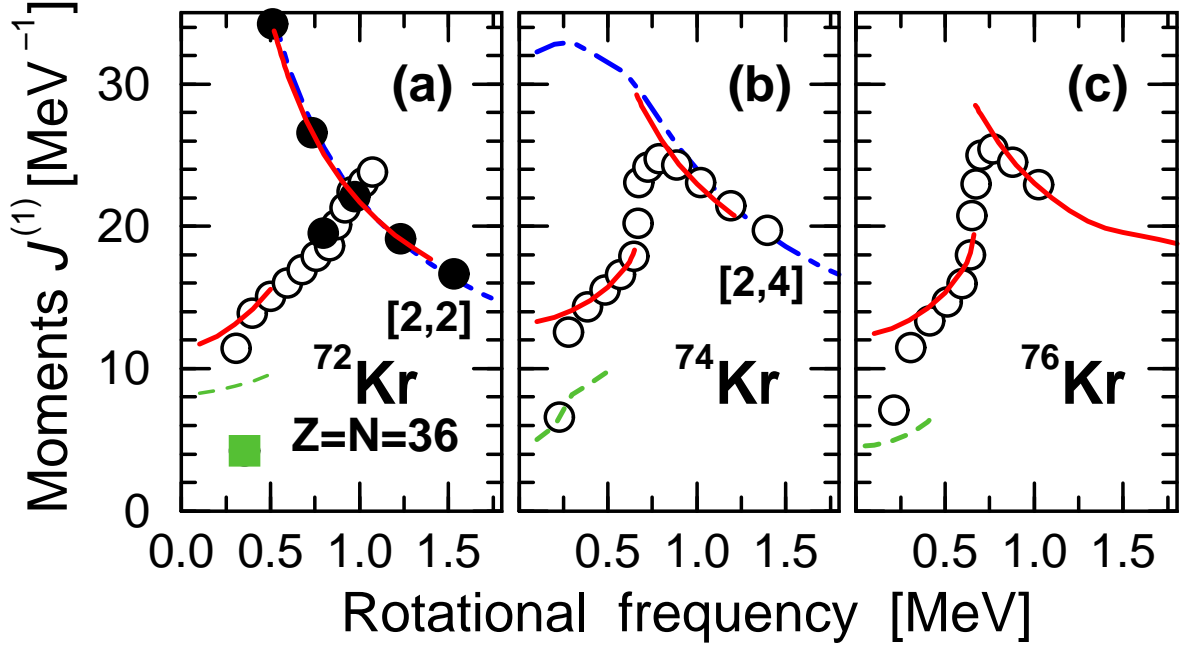}
\caption{{Experimental (symbols) and calculated (lines) kinematic 
moments of inertia in $^{72,74,76}$Kr. The results of the CRHB+LN 
calculations for the yrast configurations in the near-prolate and the 
near-oblate 
(only at low $\omega$) minima are shown. In addition, the results of the 
CRMF calculations for the lowest  configuration in  the prolate 
minimum are displayed. Bands A, B and C in $^{72}$Kr are shown by solid 
circles, open circles and solid squares, respectively. The
$J^{(1)}$ value corresponding to the linking transition between bands A and 
B with energy 1588 keV is included in the curve of band A in order to 
visualise the backbending. The experimental data are taken from 
Refs.\ \protect\cite{68Se72Kr} ($^{72}$Kr), 
\protect\cite{74Kr78Sr82Zr} ($^{74}$Kr) 
and \protect\cite{76Kr} ($^{76}$Kr).}
Solid and long-dashed lines are used for the CRHB+LN results in 
near-prolate and near-oblate minima, respectively. The CRMF results
for unpaired analogs of the S-bands are shown by dot-dashed lines.}
\label{kr-pair-j1}
\end{figure}

\begin{figure}
\epsfxsize 12.0cm
\epsfbox{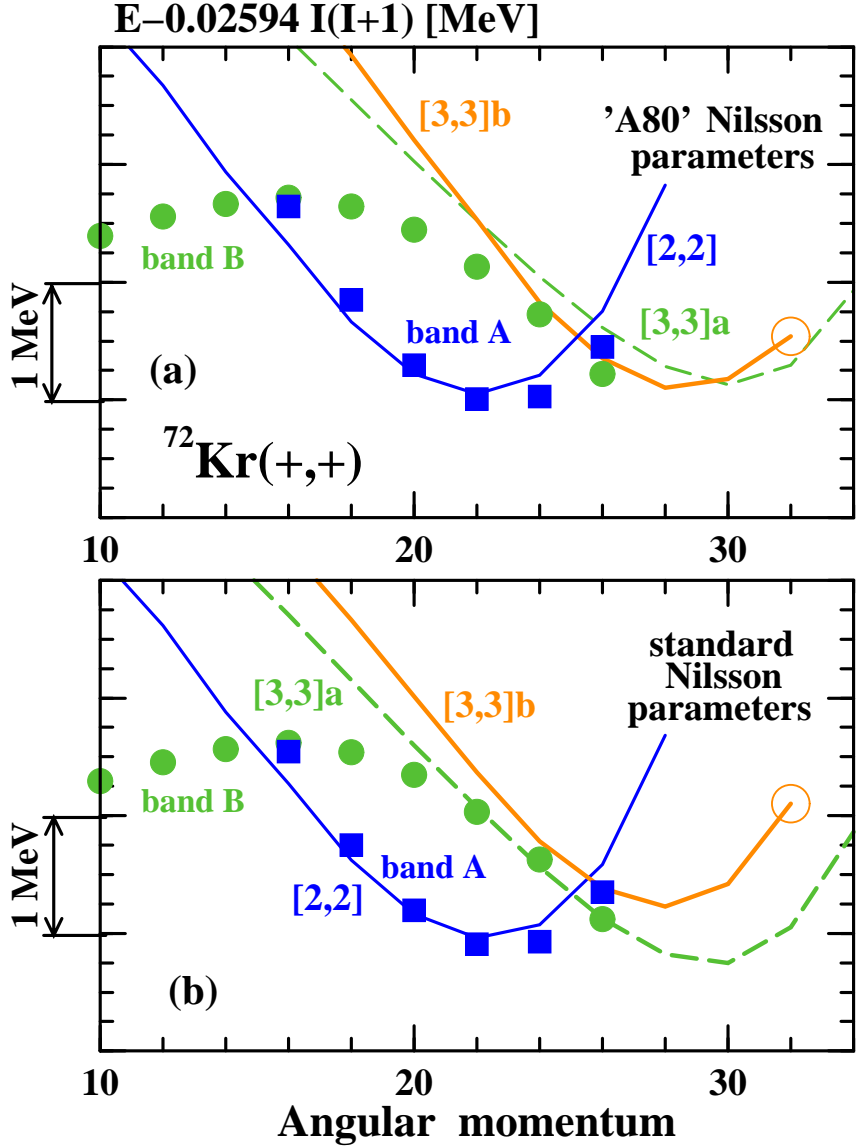}
\caption{Calculated yrast and near-yrast collective configurations 
with $\pi=+, \alpha=0$ versus observed bands A and B in $^{72}$Kr. 
The excitation energies are given relative to a rigid rotor reference 
$E_{RLD}=0.02594 I(I+1)$ MeV. The CNS calculations were performed
with the 'A80' (panel (a)) and standard (panel (b)) Nilsson 
parameters. The experimental data are taken from Ref.\ 
\protect\cite{68Se72Kr}.}
\label{kr72-eld-a80-stan}
\end{figure}

\begin{figure}
\epsfxsize 16.0cm
\epsfbox{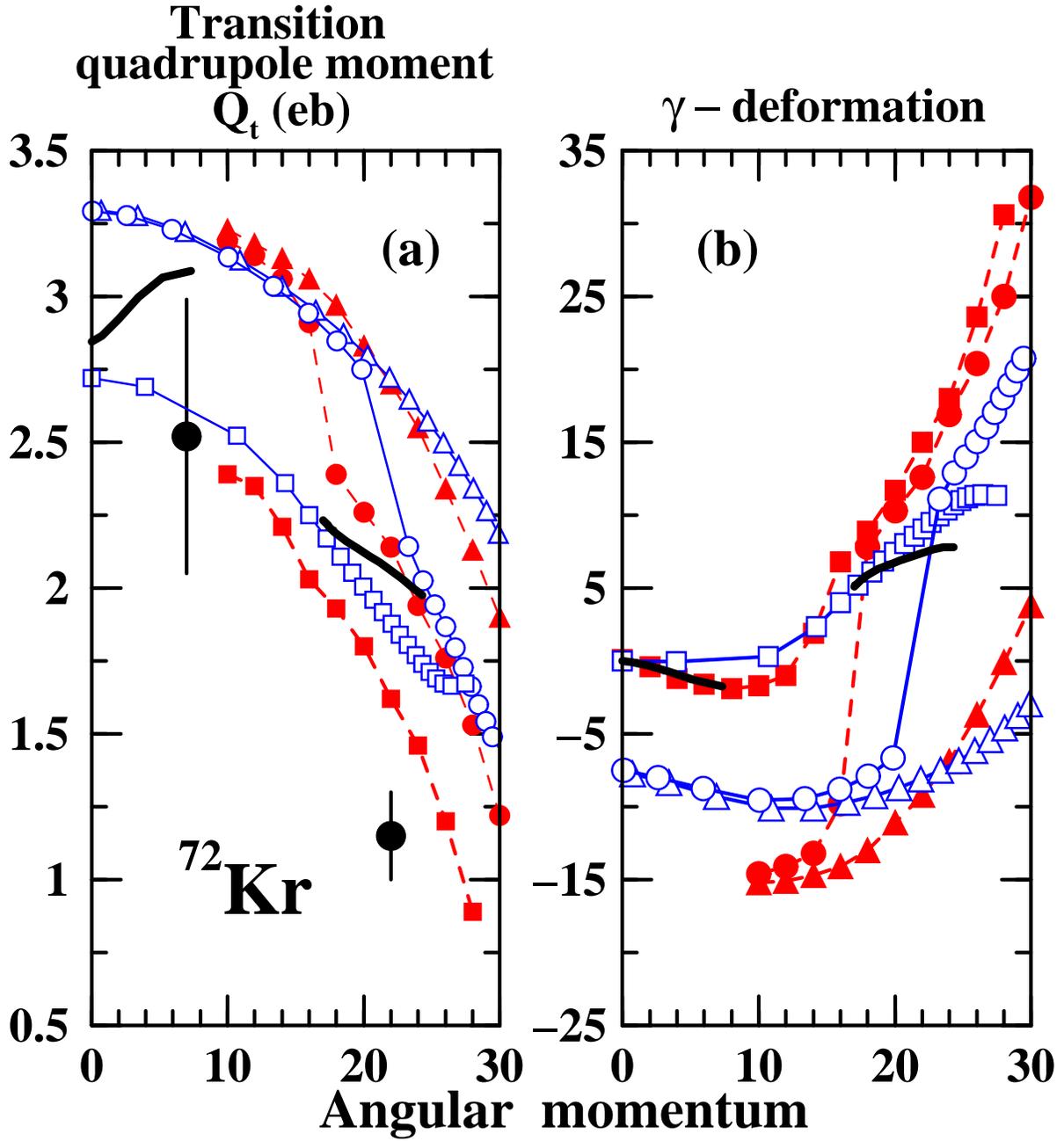}
\caption{Transition quadrupole moments $Q_t$ (panel (a)) and
$\gamma$-deformations (panel (b)) of the calculated configurations 
in $^{72}$Kr. The results of the CNS and CRMF calculations are shown 
by the solid and open symbols, respectively. The results of the CRHB+LN 
calculations for the ground state band in the prolate minimum are shown 
by thick solid lines. Squares, triangles and circles are used for the [2,2], 
[3,3]a and [3,3]b configurations, respectively. The experimental 
data are taken from Refs.\ \protect\cite{72Kr-Qt,68Se72Kr}.}
\label{kr72-def}
\end{figure}

The [2,2] and [3,3] configurations are the lowest  $(\pi=+,\alpha=0)$ states 
according to the CNS and CRMF calculations (Fig.\ \ref{kr72-eld-a80-stan}). The 
transition quadrupole moments $Q_t$ and the equilibrium deformations of these 
configurations differ considerably (Fig.\ \ref{kr72-def}). We assign the configuration 
[2,2] to band A. Fig.\ \ref{kr72-eld-a80-stan} demonstrates that the $(E-E_{RLD})$ 
curves of band A and band B (at $I\geq 20\hbar$) and their relative energies are well 
reproduced by the [2,2] and [3,3] configurations in the CNS calculations.
The same holds for the CRMF, which is not shown here since the results are too 
similar to the CNS. The kinematic moment of inertia of band A above the band 
crossing is excellently described by the CRMF and CRHB+LN calculations 
(Fig.\ \ref{kr-pair-j1}a). The low spin part of band B is interpreted
as the g-band at  prolate shape. It is well described
by the CRHB+LN calculations in the prolate minimum.  Band C is the ground 
state band in the oblate minimum. In the CRHB+LN calculations for $I=0\hbar$, 
the oblate minimum is 1.15 MeV
below the prolate one, while 
in experiment it is $\approx 0.6$ MeV \cite{68Se72Kr} lower. The prolate minimum
 takes over at $I=4\hbar$, because it has a larger moment of inertia.

  In experiment, the  g-band (band B: $I= 4-14\hbar$) is crossed by the S-band 
(band A: $I=16-26\hbar$)  at $\omega \approx  0.69$ MeV. This value is
very close to the crossing frequencies in $^{74,76}$Kr which are approximately
equal to 0.68 and 0.65 MeV, respectively. Thus recent experimental data do
not show the delay of the band crossing in the $N=Z$ system as compared with 
$N=Z+2$ system. The CRHB+LN calculations place the band crossing in $^{72}$Kr at 
$\omega \approx 0.53$ MeV (Fig.\ \ref{kr-pair-j1}a), i.e. experimental band 
crossing is by 160 keV delayed as compared with the CRHB+LN calculations, 
which reproduce very well the crossing frequencies in $^{74,76}$Kr (Fig.\ 
\ref{kr-pair-j1}b and c). However, we do not consider this discrepancy as an 
evidence for $t=0$ $np$-pairing, because comparable or even larger differences 
between calculated and experimental crossing frequencies are known for nuclei in 
which $np$-pairing does not play a role (see Sect.\ \ref{disc} for details).

 Having the structure of the g- and S-bands established, we
discuss in detail the structure of band B in the spin 
range $I=16-26\hbar$. First we focus on  high spin, where the
calculations with zero pairing apply. The CNS calculations indicate the 
presence of two closely lying [3,3] configurations (Fig.\ \ref{kr72-eld-a80-stan}), 
which are  candidates for the high-spin branch of band B. The configurations [3,3]a 
and [3,3]b are obtained from the [2,2] configuration by exciting a proton and a 
neutron from the $3_3 (\alpha=-1/2)$ and $3_3 (\alpha=+1/2)$ orbitals into second 
$g_{9/2} (\alpha=+1/2)$ orbital, respectively. The results of the CRMF calculations 
for the [2,2] and [3,3]a,b configurations are very similar to those shown in Fig.\ 
\ref{kr72-eld-a80-stan}a. The relative energies of the [2,2] and [3,3]a,b 
configurations 
depend on the energy gap between the $3_3$ and $g_{9/2}$ orbitals, which, as we 
know from $^{73,74}$Kr (see Sect.\ \ref{74Kr-unpair},\ref{Kr73-sect}), is  
well described by the CNS and CRMF calculations. Thus, the fact that for  
$I\geq 20\hbar$ the experimental energy difference between the top branch of 
band B and band A comes very close to the calculated one strongly supports the 
interpretation of top branch of band B as the [3,3] configuration. 
The details of the interpretation are, however, model
dependent. The CNS calculations with the 'A80' set and the CRMF 
calculations suggest that the  top branch of band B may be the envelope 
of the [3,3]a and [3,3]b configurations, whereas the CNS calculations with 
the standard Nilsson parameters suggests the [3,3]a configuration.
In the former case the irregularities seen in $J^{(2)}$ of the band 
B at $\omega \geq 0.8$ MeV (see Fig.\ 2 in Ref.\ \cite{Fisher}) can be explained 
as due to the crossing of the [3,3]a and [3,3]b configurations.

\begin{figure}
\epsfxsize 16.0cm
\epsfbox{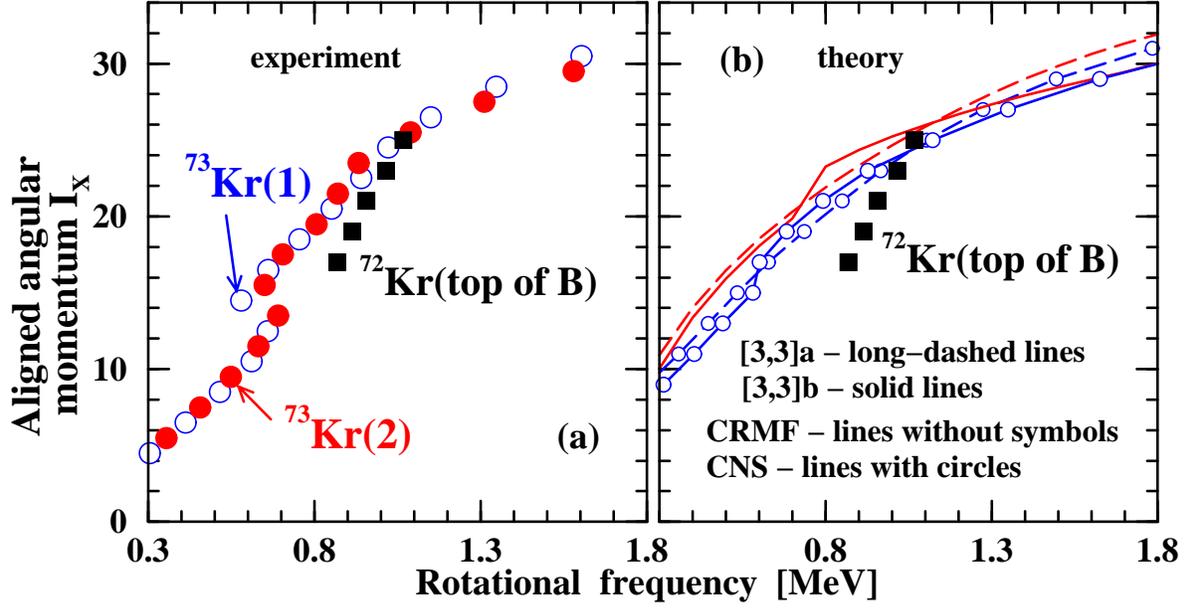}
\caption{Angular momentum  $I_x$ aligned with  the rotation axis 
as a function of rotational frequency. Panel (a) compares 
bands 1 and 2 in $^{73}$Kr with the top branch of band B in $^{72}$Kr.
Panel (b) compares the top branch of band B with the results of the CNS
(standard Nilsson parameters) and CRMF calculations. }
\label{iw-kr72-bandB-top}
\end{figure}

  Further insight into the structure of the top branch of band B can be 
obtained by comparing it with negative parity bands 1 and 2 in $^{73}$Kr 
\cite{73Kr} which are based on the [3,3] configurations (see Fig.\ 
\ref{iw-kr72-bandB-top}a). These configurations well account for the properties 
of bands 1 and 2 in $^{73}$Kr at $\omega \geq 1.0$ MeV (Fig.\ 7 in Ref.\ 
\cite{73Kr}), while the paired analogs of these configurations 
(three-quasiparticle configurations $[\pi(3)g_{9/2} \otimes \nu g_{9/2}]$) 
provide very good description of these bands within the TRS model also below a 
gradual alignment seen at $\omega \sim 1.0$ MeV (Fig.\ 8 in Ref.\ \cite{73Kr}).
As compared with the [3,3] configurations in $^{73}$Kr, the [3,3] configurations 
in $^{72}$Kr have an additional neutron hole in $3_3(\alpha=\pm 1/2)$. The unpaired 
configuration [3,3]  assigned to band B corresponds the four-quasiparticle 
$[\pi (3) g_{9/2} \otimes \nu (3) g_{9/2}]$ configuration. Unfortunately, we are 
not yet able to study this configuration within the CRHB+LN formalism.  

Comparing  the [3,3] 
configurations and the top branch of band B (Fig.\ \ref{iw-kr72-bandB-top}b) 
suggests that at frequencies higher than the  observed 
ones the aligned angular momentum of the top branch of band B cannot be 
built in the same way as at lower frequencies, because it will 
exceed the value allowed by the [3,3] configurations. Thus  the slope of 
the $I_x$ versus $\omega$ curve is expected to change around $\omega \sim 1.1$ 
MeV. This is similar to what is seen at $\omega \sim 1.0$ MeV in the bands 1 and 
2  of $^{73}$Kr. An extension of the band B to higher spin is required in 
order to check that.

 The calculated deformation parameters of the discussed configurations are shown in 
Fig.\ \ref{kr72-def}. While some differences between the CNS and CRMF results 
exist, the general features  are similar. All the configurations show a gradual 
decrease of the transition quadrupole moments $Q_t$ (Fig.\ \ref{kr72-def}a), which 
is caused by a decrease of $\beta$ and an increase of $\gamma$. Both in the CNS and 
CRMF approaches, the [3,3]b configuration jumps from negative to positive $\gamma$ 
values at $I\sim 18\hbar$ (Fig.\ \ref{kr72-def}c). If this configuration is assigned
to the top branch of band B, the jump may cause the irregularities of dynamic moment 
of inertia seen in experiment (see Fig.\ 2 in Ref.\ \cite{Fisher}).   
The low spin branch of the CRHB+LN results gives the  
$Q_t$ values for the g-band, which are close to experimental value \cite{72Kr-Qt}.
The high spin branch  of the CRHB+LN results  gives the  
$Q_t$ values for band A. As seen, 
neither of the calculations (CRHB+LN, CNS and CRMF for the [2,2] configuration) 
can explain low experimental value of $Q_t$ for the $I=22\hbar$ state of band 
A  reported in Ref.\ \cite{68Se72Kr}.
New measurements of transition quadrupole 
moments in $^{72}$Kr are needed in order to confirm or reject this discrepancy.

 According to the CNS calculations, there is a non-collective aligned state with 
[2,2] configuration and $I=16^+$, which lies $\approx 1.2$ MeV below the $I=16^+$ 
state of band A. It might be that the third observed state with $I=16^+$ at 
$E=8529$ keV (see Fig.\ 1 in Ref.\ \cite{68Se72Kr}) is this aligned state.
However, in experiment it
lies only 218 keV below the $I=16^+$ state of band A.

\subsection{$^{76}$Sr}
\label{76Sr-sect}

\begin{figure}
\epsfxsize 9.0cm
\epsfbox{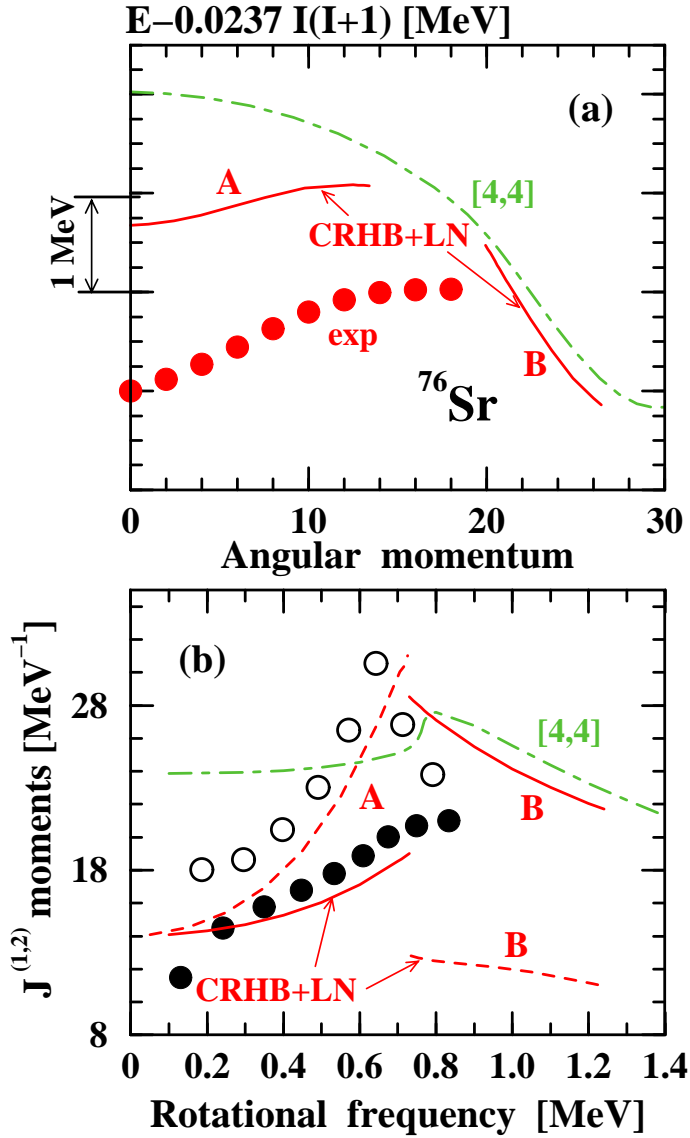}
\caption{Panel (a): Excitation energies of the ground state band in $^{76}$Sr 
(circles) compared with the CRHB+LN (solid lines) and 
the CRMF (dot-dashed line) calculations. The energies are given with respect to  
the rigid rotor reference $E_{RLD}=0.0237\, I(I+1)$ MeV. Panel (b): Kinematic 
(solid symbols) and dynamic (open symbols) moments of inertia of the 
ground state band in $^{76}$Sr compared with the CRHB+LN and CRMF results.
Solid and dashed lines are used for kinematic and dynamic moments of
inertia obtained in the CRHB+LN calculations. The dot-dashed line shows 
the kinematic moment of
inertia obtained in the CRMF calculations.
The experimental data are taken from Ref.\ \protect\cite{Fisher}.}
\label{sr76}
\end{figure}

In CRMF and CNS calculations without pairing the ($\pi=+,\alpha=0$) yrast line 
in the spin range $I=0-30\hbar$ is dominated by the collective near-prolate 
[4,4] configuration (Fig.\ \ref{sr76}a). This is in contrast to neighbouring 
$^{78}$Sr and $^{80}$Zr nuclei (Sects.\ \ref{78Sr-sect} and \ref{80Zr-sect}), 
in which the low-spin near-prolate ground state band is crossed at 
$I\sim 15-20\hbar$ by the band with negative $\gamma$-deformation. This difference 
is due to the large $N=Z=38$ 
deformed shell gap at prolate shape (see, for example, the Nilsson diagram in 
Fig.\ 8 of Ref.\ 
\cite{74Kr78Sr82Zr}). In the CNS calculations, the [4,4] configuration shows 
a gradual decrease of the transition quadrupole moment from 
$Q_t \approx 3.3$ $e$b ($\varepsilon_2 \approx 0.34$, $\gamma=-3^{\circ}$)  at 
$I=16\hbar$ down to $Q_t \approx 2.7$ $e$b ($\varepsilon_2 \approx 0.32$, 
$\gamma \approx -1^{\circ}$)  at $I=26\hbar$.

As seen in Fig.\ \ref{sr76}, most of the yrast line is well reproduced by the 
CRHB+LN calculations. Branches A and B shown in this figure are the lowest in 
energy solutions before and after band crossing. The calculation gives a sharp 
band crossing at $\omega \approx 0.73$ MeV, whereas  the experiment
shows a more gradual alignment. This discrepancy may be attributed
to the cranking approximation, which is not reliable in the region of the 
band crossing. The calculated crossing frequency is about 90 keV larger
than the experimental one. The deformation of branch A is almost 
spin-independent corresponding to $Q_t \approx 3.6$ $e$b and 
$\gamma \approx 0^{\circ}$. The branch B is a paired analog of the [4,4] 
configuration with $Q_t \approx 2.75$ $e$b and $\gamma=-5^{\circ}$ at $I=20\hbar$ and 
$Q_t \approx 2.6$ $e$b and $\gamma=-4^{\circ}$ at $I=26\hbar$.
Somewhat larger $Q_t$ values ($Q_t \approx 3.16$ $e$b at $I=20\hbar$
and $Q_t \approx 2.8$ $e$b at $I=26\hbar$) and somewhat smaller
$\gamma$-deformations ($\gamma \approx -2.5^{\circ}$) have been
obtained for the [4,4] configuration in the CRMF calculations.

\begin{figure}
\epsfxsize 10.0cm
\epsfbox{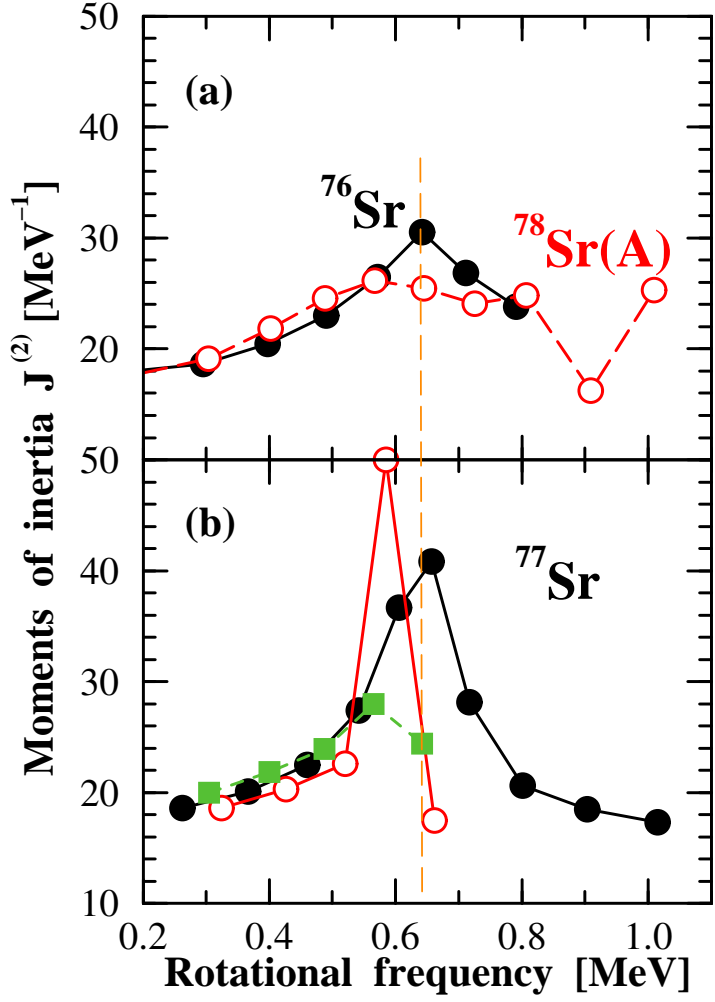}
\caption{Dynamic moments of inertia of the ground state bands in $^{76,78}$Sr (panel 
(a)) and of the one-quasiparticle bands in $^{77}$Sr (panel (b)). 
The  data are 
taken from Refs.\ \protect\cite{Fisher,74Kr78Sr82Zr,Sr77}. In the bottom panel, solid 
and open symbols are used for the $\alpha=1/2$ and $\alpha=-1/2$ bands, respectively. 
Positive and negative parity bands are denoted by the circles and squares, respectively.
The vertical long-dashed line indicates the crossing frequency in the ground state band 
of $^{76}$Sr.}
\label{j2-sr-all}
\end{figure}

  The dynamic moments of inertia of the positive parity bands in 
$^{76,78}$Sr and one-quasiparticle bands in $^{77}$Sr are 
compared in Fig.\ \ref{j2-sr-all}.  The band crossing in the $N=Z$ 
nucleus $^{76}$Sr occurs at about the same frequency as for the 
$(\pi=+,\alpha=\pm 1/2)$  and $(\pi=-,\alpha=+1/2)$ bands in $^{77}$Sr. 
The small differences may be caused by 
polarizations of the deformed potential and 
of $t=1$ pair field induced by the additional neutron in 
$^{77}$Sr. The comparison with the yrast sequence in $^{78}$Sr is 
inconclusive, because no clear bandcrossing can be seen (see Sect.\ 
\ref{78Sr-sect} for detail).

\subsection{$^{80}Zr$}
\label{80Zr-sect}

\begin{figure}
\epsfxsize 10.0cm
\epsfbox{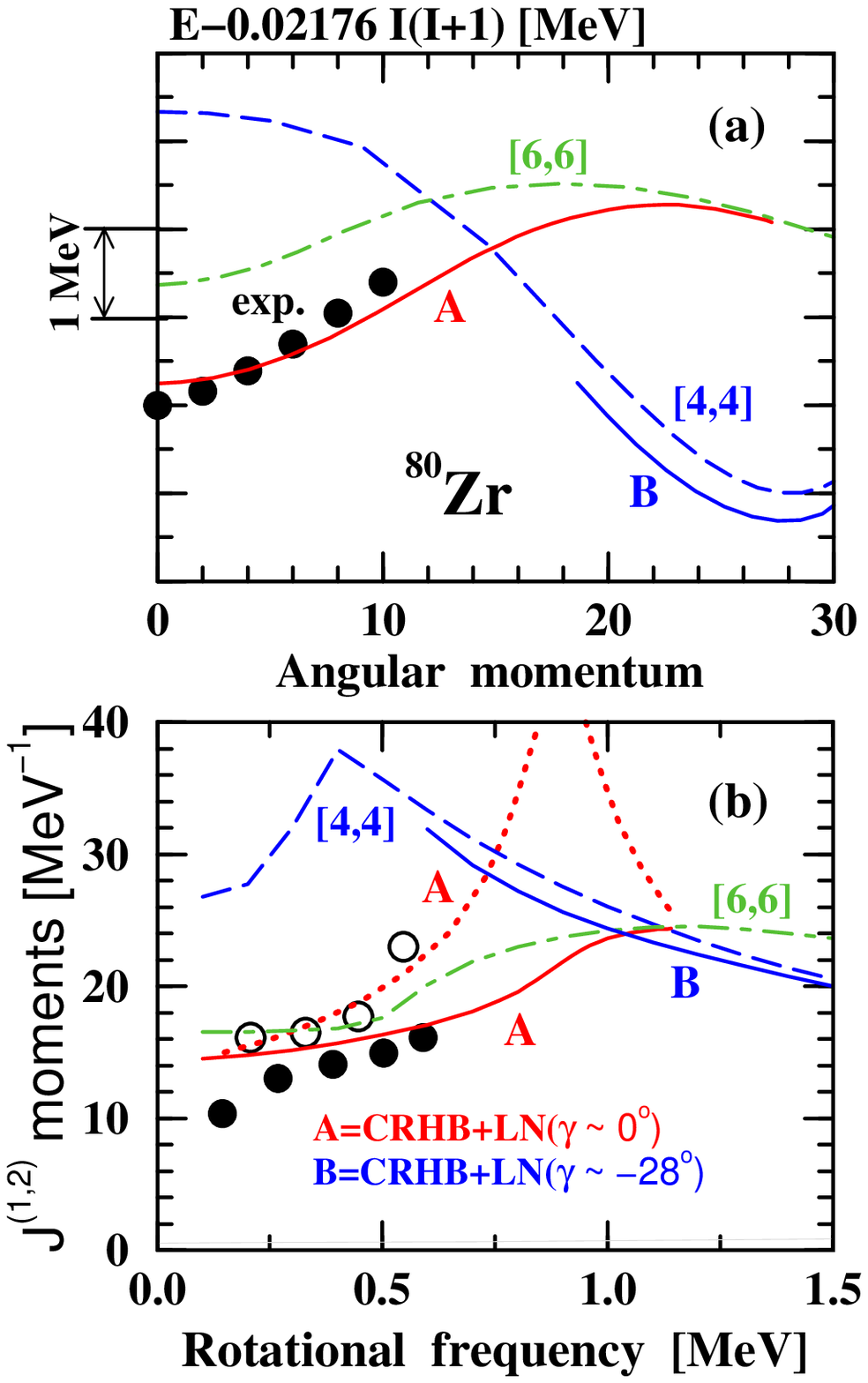}
\caption{Panel (a): Excitation energies of the ground state band in $^{80}$Zr 
(unlinked symbols) compared with the results of the CRHB+LN (solid lines) and 
the CRMF (dashed line) calculations. The energies are given  with respect of 
the rigid rotor reference $E_{RLD}=0.02176 I(I+1)$ MeV. Panel (b): Kinematic 
(solid symbols) and dynamic (open linked symbols) moments of inertia 
of ground state band in $^{80}$Zr compared with the CRHB+LN and 
CRMF results. The experimental data are taken from Ref.\ \protect\cite{Fisher}. 
The dotted line is used for the dynamic moment of inertia of configuration A, 
while other lines are used for the kinematic moments of inertia.}
\label{zr80}
\end{figure}

 According to the unpaired CRMF and CNS calculations, 
the near-prolate [6,6] configuration forms the $(\pi=+,\alpha=0)$ 
yrast line at low spin. At $I\approx 10\hbar$ it 
is crossed by the [4,4] configuration with $\gamma \sim -30^{\circ}$ (Fig.\ \ref{zr80}a),
which is the energetically favored  collective configuration in the spin range 
$I=10-28\hbar$. The paired analogs in the CRHB+LN calculations are the configurations 
denoted by A and B, respectively. Branch A is in good agreement with the experimental 
energies and moments of inertia (Fig.\ \ref{zr80}). The experimentally 
observed spins remain 
below the predicted band crossing. These two collective configurations dominate the 
yrast line. According to the CNS calculations, 
only at spins $I=16, 22, 24\hbar$ (and at $I=26\hbar$  
with the standard Nilsson parameters), the aligned or weakly collective ($I=22\hbar$) 
states of the [2,2] and [3,3] configurations are lower in energy than the [4,4] 
configuration.

  In all unpaired calculations,  the [6,6] configuration has
$Q_t \approx 3.9$ $e$b, $\varepsilon_2 \approx 0.38$ and 
$\gamma \approx -1^{\circ}$, which stay
nearly constant in the spin range $I=0\hbar-18\hbar$.
In the CNS calculation 
with the 'A80' Nilsson parameters, the [4,4] configuration 
shows a gradual decrease of the quadrupole moment from $Q_t=3.45$ $e$b 
($\varepsilon_2=0.31$, $\gamma=-34^{\circ}$)  at $I=10\hbar$  to $Q_t=2.39$ $e$b 
($\varepsilon_2=0.22$, $\gamma=-25^{\circ}$)  at $I=28\hbar$.
For the CNS calculations with the standard 
Nilsson parameters  we find a decrease from  $Q_t=3.66$ $e$b ($\varepsilon_2=0.32$, 
$\gamma=-37^{\circ}$)  at $I=10\hbar$  to $Q_t=2.88$ $e$b ($\varepsilon_2=0.26$, 
$\gamma=-33^{\circ}$)  at $I=28\hbar$.  In the CRMF calculations for this 
configurations $Q_t$ drops from 
$2.9$ $e$b ($\gamma=-24^{\circ}$) to $2.1$ $e$b ($\gamma=-24^{\circ}$) in the same 
spin range. The deformation of the [4,4]
configuration depends stronger on the model and the parametrization.
 The influence of  pairing on the deformation  of the configurations 
A and B (paired analogs of the [6,6] and [4,4]
configurations) is rather marginal, while the excitation energies, and, consequently,
the moments of inertia change stronger (see Fig.\ \ref{zr80}).


\section{Is there evidence for an isoscalar $np$-pair field?}
\label{disc}

In a number of publications it has been suggested that rotational properties 
of the $N\approx Z$ nuclei can provide evidence for the presence of a $t=0$ 
$np$ pair field. However, the reasoning often ignored the considerable $\beta$- 
and  $\gamma$-softness of the nuclei in the mass region of interest. 
In this section we are going to discuss if the comparison of the data available 
at present with our calculations, which assume a realistic (not only monopole) 
$t=1$ pair field and take the shape changes into account, suggests any evidence 
for the existence of a $t=0$ $np$ pair field. The size of the moment of inertia, the 
frequencies at which the pairs of particles align their angular momentum (band 
crossing frequencies), and unexpected mixing of configurations with a different 
number of quasiparticles have been discussed in the literature as possible
indicators of $np$-pairing in rotating $N\approx Z$ nuclei.

\subsection{Moments of inertia}
\label{momi}

Since $t=0$ pairs carry angular 
momentum, a $t=0$ $np-$pair field is expected to increase the moment of 
inertia \cite{SatW.97,MMNM.82,SatW.00,G.01}. In contrast to the $t=1$ 
pair field, which is suppressed by the Coriolis anti-pairing (CAP) effect, 
$t=0$ $np-$pairing is favored by rotation. Thus, at large angular momentum, 
where the $t=1$ field is destroyed, a substantial difference between 
experimental and calculated moments of inertia may indicate the presence 
of the $t=0$ $np-$pair field.

The moments of inertia in the g-bands of all $N=Z$ nuclei are well
reproduced by our calculations.  In particular this is true for 
$^{80}$Zr. For this nucleus Ref. \cite{G.01} carried out Cranked Hartree-Fock 
Bogoliubov calculation that included both $t=0$ and $t=1$ $np$-pair fields. 
A transition from dominating 
$t=1$ pairing to dominating $t=0$ pairing was found for the yrast line, 
which contained a spin  region where both pair fields coexist. However,
the calculated  kinematic moment of inertia increased from $J^{(1)}\approx 10$ 
MeV$^{-1}$ at $\omega \sim 0.2$ MeV 
 to $J^{(1)}\approx 40$ MeV$^{-1}$ at $\omega \sim 0.3$ MeV (see Fig.\ 
12 in Ref.\ \cite{G.01}), which is in discrepancy with experiment.

  Our calculations as well as the ones in Ref.\ \cite{SatW.00} indicate that 
after first proton and neutron paired band crossings the static $t=1$ pairing
correlations are essentially gone. In this regime of fast rotation, the 
calculations without pairing provide very good description of the rotational 
properties of different types of bands (terminating, superdeformed etc.) in 
different regions of the periodic table (see Refs.\ \cite{PhysRep,A150,A60}).
In accordance with this general observation, the experimental  moments of 
inertia in the $N=Z$ nuclei (see present manuscript for $^{68}$Se, $^{72}$Kr,
$^{76}$Se and $^{80}$Zr nuclei and Refs.\ \cite{Rb74,Zn60SD,A60,Br70} for
$^{58}$Cu, $^{60}$Zn, $^{70}$Br and $^{74}$Rb nuclei) and in the $N=Z+1$ 
nuclei (see Refs.\ \cite{73Kr} [$^{73}$Kr] and \cite{Cu59} [$^{59}$Cu])
above the first neutron and/or proton paired band crossings are well 
reproduced by the unpaired CNS and CRMF calculations, where it turned out 
to be important that the response of the nuclear shape to rotation was properly 
taken into account. No systematic underestimate of the moments of inertia, 
which could be taken as an evidence for a $t=0$ $np$-pair field, could be identified.

\subsection{Delayed band crossings}

    A delay of the first band crossing in the ground state band of an even-even 
$N=Z$ system has been widely discussed as an evidence for $t=0$ pair correlations 
\cite{SW.00,kr72t0}. HFB calculations \cite{SW.00} and shell model calculations 
\cite{kr72t0} in a $f_{7/2}$ subshell at fixed deformation indicate that the 
increase of the value of the $t=0$ $np-$pair strength results in a delay of the 
crossing frequency in the ground state band of $N=Z$ even-even nuclei. 
CSM investigations \cite{FS.99-NP,KZ.98,FS.99} show that such a delay can also 
be caused by the $t=1$ $np-$pairing. 

  As discussed in Sect.\ \ref{Kr72-sect}, the experimental situation in 
$^{72}$Kr turned out to be more complex than originally anticipated. According to 
recent data the crossing of the g- and S-bands takes place at approximately the 
same frequency as in $^{74,76}$Kr. No delay of band crossing is seen in $^{68}$Se 
as compared with $^{70}$Se (see Ref.\ \cite{68Se72Kr}).
As discussed in Sects.\ \ref{76Sr-sect} 
and \ref{78Sr-sect}, the comparison of the experimental crossing frequencies 
in $^{76}$Sr and $^{78}$Sr is not conclusive. There are no data 
permitting a comparison for $^{80}$Zr. 
Our CRHB+LN calculations give
a crossing frequency that is close to experimental data. Only for $^{72}$Kr 
the calculations underestimate it by 160 keV.

 A number of unexplained delays of the band 
crossing frequency as compared with the predictions of the CSM 
models has been observed in regions away from the $N=Z$ line
(see below), where they cannot be attributed to the $np-$pairing. 
The polarization effects at the band crossing (deformation changes, 
current changes etc.) are dramatic in the soft $A=60-80$ nuclei. 
With this in mind we conclude that no systematic delay of the 
crossing between the g-band and the doubly aligned S-band as 
compared to our CRHB+LN calculations has been found. Previous studies 
that correlate the delay of the band crossing frequencies 
with the effects of the $np-$pairing should be treated with considerable 
caution. They should be verified in the cranked mean field models which 
take these polarization effects into account in a more self-consistent 
way.

Let us mention two examples of delayed band crossings in nuclei, where
$np-$pairing is not expected to play a role.
The $\nu h_{11/2}$ crossings in the ground state bands of the 
even-even Nd $(Z=60)$ and Ce $(Z=58)$ nuclei with $A\sim 130$ 
(see Ref.\ \cite{128Nd} and references therein) are considerably 
delayed, where the maximum of this delay is situated around $N=70$ and 
$Z=58-60$. While the CSM model with the deformations taken from 
the TRS calculations predicts more or less constant value of 
crossing frequency as a function of neutron number, the experiment 
shows large variations in crossing frequencies being in the case 
of $^{132}$Nd in close agreement with calculations, while in other 
cases exceeding the CSM predictions by up to 55 \%. 
Similar delays are also seen in odd Nd and Ce nuclei and odd and 
odd-odd La $(Z=57)$ and Pr $(Z=59)$ nuclei \cite{128Nd}. In  
contrast, the $\pi h_{11/2}$ crossings are well reproduced by the 
CSM. Although predicted triaxial softness of heaviest nuclei may be 
responsible for this delay, no clear supporting evidence for that 
exists now \cite{128Nd}.
Delays of band crossings were observed also in the ground state 
bands of some rare-earth nuclei with $A\approx 180$. For example, the 
crossing in the ground state band is delayed in $^{180,182}$Hf nuclei
as compared with lower mass isotopes \cite{C-priv.02}. It was suggested 
in the framework of the projected shell model that such delays are due 
to quadrupole pairing \cite{SWF.94}. However, this is not supported by 
the CRHB+LN calculations which employ the Gogny force in pairing channel, 
and, thus, include all multipole interactions. For example, these 
calculations reproduce well the band crossing
frequencies in $^{172,178}$Hf, but fail to reproduce the delay of the band 
crossing  in $^{180}$Hf.

\subsection{Mixing of [2,2] and [3,3] configurations }

    Due to the Pauli principle, 
the $t=1$ pair field cannot scatter $np$ pairs in identical orbitals.
Only $t=0$  $np$ pairs can be 
in identical states and be scattered.
Therefore, strong mixing of configurations which differ 
by a change of such $t=0$ $np$ pair may be an evidence for $t=0$ pair 
correlations. However, it has to be mentioned, that this is not necessarily 
evidence for the presence of a $t=0$ pair field, because a finite matrix 
element of the residual interaction can cause a substantial mixing of two 
configurations with close energies.   

     This type of configuration mixing has first been discussed for $^{73}$Kr
\cite{73Kr}. It was speculated that in bands 2 and 3 at intermediate spin 
 the configuration changes from $(\pi g_{9/2}^2 \otimes \nu (fp) g_{9/2}^2)$   to 
$(\pi (fp) g_{9/2} \otimes \nu (fp) g_{9/2}^2)$, which corresponds to
a transfer of a neutron-proton pair from the $pf$ orbital into
the $g_{9/2}$ orbital.

In $^{70}$Br and $^{72}$Kr, the [2,2] and [3,3] configurations are 
located very close in energy (see Fig.\ 14 in Ref.\cite{Br70} and 
Fig.\  \ref{kr72-eld-a80-stan} in the present manuscript). If the
$t=0$ $np$-pairing is present, these configurations are expected 
to be mixed.  A mixing represents the scattering of 
a proton and neutron on identical negative parity $N=3$ orbitals into 
identical $g_{9/2}$ orbitals, and vise versa. Such a pair has an 
isospin $t=0$, since the proton and neutron are in the same space-spin 
state. It was speculated in Sect.\ VID of Ref.\ \cite{Br70} that the
observation of one smooth rotational band HD1 in $^{70}$Br at high 
spin instead of the calculated distinct crossing between the [2,2] 
and [3,3] configurations at $I \sim 18\hbar$ may point to a such 
mixing. 

More detailed experimental investigation of spectra in $^{70}$Br in 
the spin range $I=10-30\hbar$ (and, in particular, the observation of
the bands associated with the [2,2] configurations) is needed to clarify 
the situation in this nucleus. In particular, a determination of $Q_t$
for band HB1 via lifetime measurements could clarify the situation,
since the calculated $Q_t$ is twice as large for the 
$[3,3](\gamma \sim -20^{\circ})$ configuration as for the [2,2] 
configurations (see insert in Fig.\ \ref{br70-defpath}).

In $^{72}$Kr, the high spin  
branch of band B is assigned to the [3,3] configuration. 
 Band B is the g-band 
at low spin, which is a superposition of the configurations [2,2] and 
[4,4]. This means that band B changes from this structure at low spin to 
[3,3] at medium/high spin. This change 
proceeds in a reasonably smooth way (Fig.\ \ref{kr72-eld-a80-stan}). It 
is possible that the peak in the dynamical moment of  inertia seen at 
0.85 MeV (see Fig.\ 2 in Ref.\ \cite{Fisher}) reflects this change. The 
smooth transition from the [2,2]+[4,4] structure at low spin to [3,3] at 
high spin would indicate a scattering of $t=0$ $np$ pairs. 
However, this explanation implies that there should also be such 
scattering between bands B and A, which are assigned to [3,3] and [2,2] 
configurations, respectively. The scattering should shift the levels of 
band A and lead to the  modifications of the dynamic moment of inertia 
at these frequencies, which is, however, smooth in experiment. Also the 
small distance of the $I=16^+$ states of bands A and B speaks against 
a strong interaction between the two bands. However, for a final judgment
one has to carry a three-band (g, [2,2], [3,3]) mixing analysis.

The discussed band mixings might be interpreted as related to the scattering 
of $np$ pairs on identical single-particle states in the $N=3$  shell and the 
$g_{9/2}$ shell.  However we do not consider this as sufficient evidence for 
the presence of a $t=0$ pair field. Rather it may indicate weak dynamical $t=0$ 
pair correlations as suggested by 
the Monte Carlo shell model calculations \cite{DKLR.97} or just 
mixing of energetically close configurations by residual interaction.

\section{Conclusions}
\label{concl}

  The rotational bands in even-even $N \approx Z$ nuclei have been studied by 
means of the cranked Nilsson+Strutinsky approach, cranked relativistic mean 
field and cranked relativistic Hartree+Bogoliubov theories. Due to the spontaneous 
breaking of the isospin symmetry by the isovector pair field, its $np$ component 
needs not to be taken into account explicitly.  It is included implicitly when 
the isospin symmetry is restored by adding the isorotational energy $T(T+1)/2J_{iso}$ 
to the intrinsic energy \cite{FS.99-NP}. Since $T=1$ states lie at a substantial 
excitation energy in even-even $N=Z$ nuclei, only $T=0$ rotational bands were 
considered, which means that the isorotational energy shift is irrelevant for 
the rotational response.  The systematic investigation of the rotating even-even 
$N=Z$ nuclei with $A=68-80$ leads to the following conclusions.
 
 Coexistence of prolate and oblate shapes is found near the ground state.
For $I>2$, the yrast states have prolate shape. Their moments of inertia are
well reproduced by the calculations that take only the isovector pair
field into account. At high spin, the isovector pairing is very weak and 
calculations with zero pairing agree well with the results of treating the 
isovector pairing in the framework of the Lipkin-Nogami approximation. All 
these calculations account well for the data. This concerns the moments of 
inertia as well as the relative energies of different configurations. The level 
of agreement between  theory and experiment is comparable, sometimes
even better than for the nuclei away from the $N=Z$ line. The available 
experimental high-spin data do not require the introduction of the $t=0$ 
$np$-pairing into the models, and, thus, do not provide any evidence 
of this type of pairing.

  The delay of the crossing between the ground band and the doubly aligned 
S-band in even-even $N=Z$ nuclei has been suggested as an evidence for the 
$t=0$ \cite{SatW.97,SW.00} $np$ pair field. The new data do not show such a 
delay in $^{72}$Kr, which was considered as the most important evidence. Such 
delay is also absent in $^{68}$Se \cite{68Se72Kr}. The data on other even-even
nuclei do not permit us to compare the band crossing frequencies in $N=Z$ and 
$N=Z+2$ isotopes.  The properties of bands in paired regime and the paired 
band crossings in $^{68}$Se (Sect.\ \ref{Se68-sect}), $^{72}$Kr (Sect.\ 
\ref{Kr72-sect}), $^{76}$Sr (Sect.\ \ref{76Sr-sect}) and $^{80}$Zr (Sect.\ 
\ref{80Zr-sect}) are reasonably well described by the CRHB+LN theory (which 
implicitly takes into account the isovector $np$-pairing). The minor differences 
between experimental and calculated band crossing frequencies are within the 
expected accuracy of our theoretical tools.

  There seems to be strong mixing between some configurations that are 
related to each other by the transfer of a $t=0$ $np$ pair on identical
single particle states, which may be caused by $t=0$ $np$ pair 
correlations. However, this cannot be considered as sufficient evidence 
for the presence of a $t=0$ pair field. 

  The systematic analysis of the rotational response of even-even $N=Z$ nuclei
confirms the  picture suggested in Ref.\ \cite{FS.99-NP}. At low spin, there 
is no isoscalar $np$ pair field but a strong isovector pair field exists, 
which includes a large $np$ component, the strength of which is determined by
isospin conservation. Like in nuclei away from the $N=Z$ line, this isovector 
pair field is destroyed by rotation. In this high spin regime calculations 
without pairing describe well the data provided the drastic shape changes 
that cause among other things band termination are taken into account.

\section{Acknowledgements}

We thank Carl Svensson for providing and allowing us to use recent
unpublished experimental data on $^{74}$Kr. The work was supported by the
DoE grant DE-F05-96ER-40983.

\newpage

\newpage

\newpage

\newpage

\newpage

\newpage

\begin{figure}
\epsfxsize 16.0cm
\caption{Excitation energies of the configurations forming the yrast lines 
of 4 combinations of parity and signature in $^{64}$Ge relative to a 
rigid rotor reference $E_{RLD}=0.03157 I(I+1)$ MeV. Calculated terminating 
(aligned) states are encircled. The shorthand notation $<p_1p_2,n_1n_2>$ 
indicates the number $p_1(n_1)$ of occupied $g_{9/2}$ proton (neutron) 
orbitals and the number $p_2(n_2)$ of occupied $h_{11/2}$ proton (neutron) 
orbitals. $p_2(n_2)$ are omitted when the latter orbitals are not occupied. The wide 
line indicates the total yrast line. The same type of symbols is used for 
signature partner orbitals. Solid (open) symbols are used for $\alpha=0 (1)$ 
configurations.}
\label{Ge64-eld}
\end{figure}

\newpage

\newpage

\newpage

\newpage

\newpage

\newpage

\newpage


\begin{thebibliography}{99}

\bibitem{engel}  J.\ Engel, K.\ Langanke, and P.\ Vogel,  
Phys.\ Lett.\ B {\bf 389}, 211 (1996). 

\bibitem{EPSVD.97} J.\ Engel, S.\ Pittel, M.\ Stoitsov, P.\ Vogel,
and J.\ Dukelsky, Phys.\ Rev. C {\bf 55}, 1781 (1997).

\bibitem{DKLR.97}
D.\ J.\ Dean, S.\ E.\ Koonin, K.\ Langanke and P.\ B.\ Radha, Phys.\ Lett. 
B {\bf 399}, 1 (1997).

\bibitem{bes-rev} D.\ R.\ Bes, R.\ A.\ Broglia, O.\ Hansen, O.\ Nathan, 
Phys.\ Rep. {\bf 34}, 1 (1977). 

\bibitem{MFCC.00}  A.\ O.\ Macchiavelli, P.\ Fallon, R.\ M.\ Clark, M.\ Cromaz, 
M.\ A.\ Deleplanque, R.\ M.\ Diamond, G.\ J.\ Lane, I.\ Y.\ Lee, F.\ S.\ Stephens, 
C.\ E.\ Svensson, K.\ Vetter, and D.\ Ward, Phys.\ Lett. B {\bf 480}, 1 (2000).

%
\bibitem{alt} J.\ J{\"a}necke, Nucl.\ Phys. {\bf 73}, 97 (1965).

\bibitem{V.00} P.\ Vogel, Nucl.\ Phys. {\bf A662}, 148 (2000).

\bibitem{MFC.00} A.\ O.\ Macchiavelli, P.\ Fallon, R.\ M.\ Clark, M.\ Cromaz, 
M.\ A.\ Deleplanque, R.\ M.\ Diamond, G.\ J.\ Lane, I.\ Y.\ Lee, F.\ S.\ Stephens, 
C.\ E.\ Svensson, K.\ Vetter, and D.\ Ward, Phys.\ Rev. C {\bf 61}, 041303(R) 
(2000). 

%
\bibitem{Rb74} C.\ D.\ O'Leary, C.\ E.\ Svensson, S.\ G.\ Frauendorf,
A.\ V.\ Afanasjev, D.\ E.\ Appelbe, R.\ A.\ E.\ Austin, 
G.\ C.\ Ball, J.\ A.\ Cameron, R.\ M.\ Clark, M.\ Cromaz, 
P.\ Fallon, D.\ F.\ Hodgson, N.\ S.\ Kelsall, A.\ O.\ 
Macchiavelli, I.\ Ragnarsson, D.\ Sarantites, J.\ C.\ 
Waddington and R.\ Wadsworth, Phys.\ Rev. C {\bf 67}, 021301(R) (2003).

\bibitem{Beng85} T.\ Bengtsson and I.\ Ragnarsson, Nucl.\ Phys. 
{\bf A436}, 14 (1985).

\bibitem{A110} A.\ V.\ Afanasjev and I.\ Ragnarsson, Nucl.\ Phys. 
{\bf A591}, 387 (1995).

%
\bibitem{PhysRep}
A.\ V.\ Afanasjev, D.\ B.\ Fossan, G.\ J.\ Lane and I.\ Ragnarsson, Phys.\ Rep.
{\bf 322}, 1 (1999).

\bibitem{KR.89} W.\ Koepf and P.\ Ring, Nucl.\ Phys. {\bf A493}, 61 (1989).

\bibitem{KR.93} J.\ K{\"o}nig and P.\ Ring, Phys.\ Rev.\ Lett. {\bf 71}, 
3079 (1993).

\bibitem{A150} A.\ V.\ Afanasjev, J.\ K\"onig and P.\ Ring, Nucl.\ 
Phys. {\bf A608}, 107 (1996).

\bibitem{CRHB} A.\ V.\ Afanasjev, P.\ Ring, and J.\ K{\"o}nig,
Nucl.\ Phys. {\bf A676}, 196 (2000).

%
\bibitem{Zn60SD} C.\ E.\ Svensson, D.\ Rudolph, C.\ Baktash,
M.\ A.\ Bentley, J.\ A.\ Cameron, M.\ P.\ Carpenter, M.\ Devlin,
J.\ Eberth, S.\ Flibotte, A.\ Galindo-Uribarri, G.\ Hackman,
D.\ S.\ Haslip, R.\ V.\ F.\ Janssens, D.\ R.\ LaFosse, T.\ J.\ Lampman,
I.\ Y.\ Lee, F.\ Lerma, 
A.\ O.\ Macchiavelli, J.\ M.\ Nieminen, S.\ D.\ Paul, D.\ C.\ Radford,
P.\ Reiter, L.\ L.\ Riedinger, D.\ G.\ Sarantites, B.\ Schaly,
D.\ Seweryniak, O.\ Thelen, H.\ G.\ Thomas, J.\ C.\ Waddington,
D.\ Ward, W.\ Weintraub, J.\ N.\ Wilson, C.\ H.\ Yu,
A.\ V.\ Afanasjev, and I.\ Ragnarsson, Phys.\ Rev.\ Lett. {\bf 82}, 
3400 (1999).

%
\bibitem{A60} A.\ V.\ Afanasjev, I.\ Ragnarsson and  P.\ Ring, Phys.\ Rev. 
C {\bf 59}, 3166 (1999).

\bibitem{Cu59} C.\ Andreoiu, D.\ Rudolph, C.\ E.\ Svensson, A.\ V.\ Afanasjev, 
J.\ Dobaczewski, I.\ Ragnarsson, C.\ Baktash, J.\ Eberth, C.\ Fahlander, D.\ S.\ Haslip, 
D.\ R.\ LaFosse, S.\ D.\ Paul, D.\ G.\ Sarantites, H.\ G.\ Thomas, J.\ C.\ Waddington, 
W.\ Weintraub, J.\ N. Wilsson and C.-H.\ Yu, Phys.\ Rev. C {\bf 62}, 051301(R) (2000). 

\bibitem{J1Rare}
A.\ V.\ Afanasjev, J.\ K{\"o}nig, P.\ Ring, L.\ M.\ Robledo,
and J.\ L.\ Egido, Phys.\ Rev.\ C {\bf 62}, 054306 (2000).

%
\bibitem{A250} A.\ V.\ Afanasjev, T.\ L.\ Khoo, S.\ Frauendorf, G.\ A.\ Lalazissis, 
and I.\ Ahmad, Phys.\ Rev. C {\bf 67}, 024309 (2003).

\bibitem{A190} A.\ V.\ Afanasjev, J.\ K{\"o}nig, and P.\ Ring, Phys.\ Rev. C {\bf 60}, 
051303 (1999).
                                                                                
\bibitem{VRAL} D.\ Vretenar, A.\ V.\ Afanasjev, G.\ Lalazissis, and P.\ Ring, 
submitted to Physics Reports

\bibitem{WS.01} R.\ A.\ Wyss and W.\ Satu{\l}a, Acta Phys.\ Polonica B {\bf 32}, 2457 (2001).

\bibitem{S.03} Y.\ Sun, Eur.\ Phys.\ Jour. {\bf 20}, 133 (2004).

\bibitem{PSF.02} A.\ Petrovici, K.\ W.\ Schmid, and A.\ Faessler, Nucl.\ Phys. {\bf A 710},
 246 (2002).

\bibitem{FS.99-NP} S.\ G.\ Frauendorf and J.\ A.\ Sheikh, Nucl.\ Phys. 
{\bf A645}, 509 (1999).

\bibitem{NL3} G.\ A.\ Lalazissis, J.\ K\"onig and P.\ Ring, Phys.\ Rev. C {\bf 55}, 540 (1997).

\bibitem{D1S} J.\ F.\ Berger, M.\ Girod, and D.\ Gogny, Comp.\ Phys.\ Comm. {\bf 63}, 
365 (1991).

\bibitem{GBI.86} D.\ Galeriu, D.\ Bucurescu, and M.\ Iva\c{s}ku, J.\ Phys. G {\bf 12}, 329 
(1986).

\bibitem{NDBBR.85} W.\ Nazarewicz, J.\ Dudek, R.\ Bengtsson, T.\ Bengtsson, and I.\ Ragnarsson, 
Nucl.\ Phys. {\bf A435}, 397 (1985).

%
\bibitem{Zn62bt} C. E. Svensson,
C.\ Baktash, G.\ C.\ Ball, J.\ A.\ Cameron, M.\ Devlin, J.\ Eberth,
S.\ Flibotte, A.\ Galindo-Uribarri,  D.\ S.\ Haslip, V.\ P.\ Janzen,
D.\ R.\ LaFosse, I.\ Y.\ Lee, A.\ O.\ Macchiavelli, R.\ W.\ MacLeod,
J.\ M.\ Nieminen, S.\ D.\ Paul, D.\ C.\ Radford, L.\ L.\ Riedinger,
D.\ Rudolph, D.\ G.\ Sarantites, H.\ G.\ Thomas, J.\ C.\ Waddington,
D.\ Ward, W.\ Weintraub, J.\ N.\ Wilson, A.\ V.\ Afanasjev and 
I.\ Ragnarsson, Phys.\ Rev.\ Lett. {\bf 80}, 2558 (1998).

%
%
\bibitem{Br73}
C.\ Plettner, H.\ Schnare, R.\ Schwengner, L.\ K\"aubler, F.\ D\"onau, 
I.\ Ragnarsson, A.\ V.\ Afanasjev, A.\ Algora, G.\ de Angelis, A.\ Gadea, 
D.\ R.\ Napoli, J.\ Eberth, T.\ Steinhardt, O.\ Thelen, M.\ Hausmann, 
A.\ M\"uller, A.\ Jungclaus, K.\ P.\ Lieb, D.\ G.\ Jenkins, R.\ Wadsworth, 
A.\ N.\ Wilson and S.\ Frauendorf, Phys.\ Rev. C {\bf 62}, 014313 (2000).  

\bibitem{SatW.97} S.\ Satu{\l}a and R.\ Wyss, Phys.\ Lett. B {\bf 393}, 1 (1997).

\bibitem{84Zr} R.\ Cardona, F.\ Cristancho, S.\ L.\ Tabor, R.\ A.\ Kaye, G.\ Z.\ Solomon, 
J.\ D{\"o}ring, G.\ D.\ Johns, M.\ Devlin, F.\ Lerma, D.\ G.\ Sarantites, I.-Y.\ Lee, 
A.\ O.\ Macchiavelli, I.\ Ragnarsson, Phys.\ Rev. C {\bf 68}, 024303 (2003). 

\bibitem{Kr74} C.\ E.\ Svensson, private communication (2003).

\bibitem{76Kr} S.\ Balraj, Nucl.\ Data Sheets {\bf 74}, 63 (1995); C.\ E.\ Svensson, 
C.\ D.\ O'Leary, I.\ Ragnarsson, D.\ E.\ Appelbe, R.\ A.\ E.\ Austin, G.\ C.\ Ball, 
J.\ A.\ Cameron, R.\ M.\ Clark, M.\ Cromaz, P.\ Fallon, D.\ F.\ Hodgson, N.\ S.\ Kelsall, 
A.\ O.\ Macchiavelli, D.\ Sarantites, J.\ C.\ Waddington, R.\ Wadsworth, and D.\ Ward, 
submitted to Phys.\ Rev. C 

\bibitem{73Kr} N.\ S.\ Kelsall, S.\ M.\ Fischer, D.\ P.\ Balamuth, G.\ C.\ Ball, 
M.\ P.\ Carpenter, R.\ M.\ Clark, J.\ Durell, P.\ Fallon,  S.\ J.\ Freeman, 
P.\ A.\ Hausladen, R.\ V.\ F.\ Janssens, D.\ G.\ Jenkins, M.\ J.\ Leddy, 
C.\ J.\ Lister, A.\ O.\ Macchiavelli, D.\ G.\ Sarantites, D.\ C.\ Schmidt, 
D.\ Seweryniak, C.\ E.\ Svensson, B.\ J.\ Varley, S.\ Vincent, R.\ Wadsworth,
A.\ N.\ Wilson, A.\ V.\ Afanasjev, S.\ Frauendorf, I.\ Ragnarsson and 
R.\ Wyss, Phys.\ Rev. C {\bf 65}, 044331 (2002).

\bibitem{74Kr-Qt} A.\ Algora, G. de Angelis, F.\ Brandolini, R.\ Wyss, A.\  Gadea, 
E.\ Farnea, W.\ Gelletly, S.\ Lunardi, D.\ Bazzacco, C.\ Fahlander, A.\ Aprahamian, 
F.\ Becker, P.\ G.\ Bizzetti, A.\ Bizzetti-Sona, D.\ de Acu\~{n}a, M.\ De Poli, 
J.\ Eberth, 
D.\ Foltescu, S.\ M.\ Lenzi, T.\ Martinez, D.\ R.\ Napoli, P. Pavan, C.\ M.\ Petrache, 
C.\ Rossi Alvarez, D.\ Rudolph, B.\ Rubio, S.\ Skoda, P.\ Spolaore, R.\ Menegazzo,
H.\ G.\ Thomas, and C.\ A.\ Ur, Phys.\ Rev. C {\bf 61}, 031303(R) (2000). 

\bibitem{74Kr78Sr82Zr} D.\ Rudolph, C.\ Baktash, C.\ J.\ Gross, W.\ Satu{\l}a, 
R.\ Wyss, I.\ Birriel, M.\ Devlin, H.-Q.\ Jin, D.\ R.\ LaFosse, 
F.\ Lerma, J.\ X.\ Saladin, D.\ G.\ Sarantites, G.\ N.\ Sylvan, 
S.\ L.\ Tabor, D.\ F.\ Winchell, V.\ Q.\ Wood, and C.\ H.\ Yu,
Phys.\ Rev.\ C {\bf 56}, 98 (1997).



\bibitem{74Kr?} C.\ Chandler, P.\ H.\ Regan, C.\ J.\ Pearson, B.\ Blank, 
A.\ M.\ Bruce, W.\ N.\ Catford, N.\ Curtis, S.\ Czajkowski, W.\ Gelletly, 
R.\ Grzywacz, Z.\ Janas, M.\ Lewitowicz, C.\ Marchand, N.\ A.\ Orr, 
R.\ D.\ Page, A.\ Petrovici, A.\ T.\ Reed, M.\ G.\ Saint-Laurent, 
S.\ M.\ Vincent, R.\ Wadsworth, D.\ D.\ Warner, and J.\ S.\ Winfield,
Phys.\ Rev.\ C {\bf 56}, 2924(R) (1997).

 
\bibitem{Pingst-A30-60} A.\ V.\ Afanasjev, P.\ Ring and I.\ Ragnarsson,
Proc. Int. Workshop PINGST2000 "Selected topics on $N=Z$ nuclei", 
2000, Lund, Sweden, Eds. D.\ Rudolph and M. Hellstr{\"o}m, (2000) 
p.\ 183.

\bibitem{Fisher} S.\ M.\ Fischer, C.\ J.\ Lister, D.\ P.\ Balamuth, R.\ Bauer, 
J.\ A.\ Becker, L.\ A.\ Bernstein, M.\ P.\ Carpenter, J.\ Durell, N.\ Fotiades, 
S.\ J.\ Freeman, P.\ E.\ Garrett, P.\ A.\ Hausladen, R.\ V.\ F.\ Janssens, D.\ Jenkins, 
M.\ Leddy, J.\ Ressler, J.\ Schwartz, D.\ Svelnys, D.\ G.\ Sarantites, D.\ Seweryniak, 
B.\ J.\ Varley, and R.\ Wyss, Phys.\ Rev.\ Lett. {\bf 87}, 132501 (2001).

\bibitem{77Rb-1} A.\ Harder, F.\ D\"onau, K.\ P.\ Lieb, R.\ A.\ Cunningham, 
W.\ Gelletly, C.\ J.\ Gross, F.\ Hannachi, M.\ K.\ Kabadiyski, H.\ A.\ Roth,
D.\ Rudolph, J.\ Simpson, \"O.\ Skeppstedt, B.\ J.\ Varley, and D.\ D.\ Warner, 
Phys.\ Lett. B {\bf 374}, 277 (1996).

\bibitem{77Rb-2}
A. Harder, A. Jungclaus, M. K. Kabadiyski, D. Kast, K. P. Lieb, D. Rudolph, 
M. Weiszflog, T. D. Johnson, G. Winter, C. J. Gross, R. A. Cunningham, 
W. Gelletly, J. Simpson, D. D. Warner, I. G. Bearden, T. Shizuma, 
G. Sletten, D. Foltescu, H. A. Roth, \"O. Skeppstedt, and B. J. Varley,
Phys. Rev. C {\bf 55}, 1680 (1997).

\bibitem{Sr77} C.\ J.\ Gross, C.\ Baktash, D.\ M.\ Cullen, R.\ A.\ Cunningham, 
J.\ D.\ Garrett, W.\ Gelletly, F.\ Hannachi, A.\ Harder, M.\ K.\ Kabadiyski, 
K.\ P.\ Lieb, C.\ J.\ Lister, W.\ Nazarewicz, H.\ A.\ Roth, D.\ Rudolph, 
D.\ G.\ Sarantites, J.\ A.\ Sheikh, J.\ Simpson, \"O.\ Skeppstedt, B.\ J.\ Varley, and 
D.\ D.\ Warner, Phys.\ Rev. C {\bf 49}, R580 (1994).

\bibitem{AR-unp} A.\ V.\ Afanasjev and I.\ Ragnarsson, unpublished

\bibitem{Ge64-exp} P.\ J.\ Ennis, C.\ J.\ Lister, W.\ Gelletly, H.\ G.\ Price,
B.\ J.\ Varley, P.\ A.\ Butler, T.\ Hoare, S.\ \'{C}wiok, W.\ Nazarewicz,
Nucl.\ Phys. A 535 (1991) 392

\bibitem{68Se72Kr} S.\ M.\ Fischer, C.\ J.\ Lister, and D.\ P.\ Balamuth, Phys.\ 
Rev. C {\bf 67}, 064318 (2003).

%
\bibitem{Br70}
D.\ G.\ Jenkins, N.\ S.\ Kelsall, C.\ J.\ Lister, D.\ P.\ Balamuth, 
M.\ P.\ Carpenter, T.\ A.\ Sienko, S.\ M.\ Fischer, R.\ M.\ Clark, 
P.\ Fallon, A.\ G\"orgen, A.\ O.\ Macchiavelli, C.\ E.\ Svensson, 
R.\ Wadsworth, W.\ Reviol, D.\ G.\ Sarantites, G.\ C.\ Ball, J.\ Rikovska Stone, 
O.\ Juillet, P.\ van Isacker, A.\ V.\ Afanasjev and S.\ Frauendorf, Phys.\ Rev. C 
{\bf 65}, 064307 (2002).  

\bibitem{Se68-old} S.\ M.\ Fischer, D.\ P.\ Balamuth,  
P.\ A.\ Hausladen, C.\ J.\ Lister, M.\ P.\ Carpenter,
D.\ Seweryniak and J.\ Schwartz,
Phys.\ Rev.\ Lett. {\bf 84}, 4064 (2000).

\bibitem{SMAW.03} Y.\ Sun, Z.\ Ma, A.\ Aprahamian, and M.\ Wiescher,
preprint nucl-th/0107004

%
\bibitem{Berk-2} N.\ S.\ Kelsall, C.\ E.\ Svensson, S.\ Fischer,
D.\ E.\ Appelbe, R.\ A.\ E.\ Austin, D.\ P.\ Balamuth, 
G.\ C.\ Ball, J.\ A.\ Cameron, M.\ P.\ Carpenter, R.\ M.\ Clark,
M.\ Cromaz, M.\ A.\ Deleplanque, R.\ M.\ Diamond, J.\ L.\ Durell,
P.\ Fallon, S.\ J.\ Freeman, P.\ A.\ Hausladen, D.\ F.\ Hodgson,
R.\ V.\ F.\ Janssens, D.\ G.\ Jenkins, G.\ J.\ Lane, M.\ J.\ Leddy,
C.\ J.\ Lister, A.\ O.\ Macchiavelli, C.\ D.\ O'Leary, D.\ G.\ Sarantites,
F.\ S.\ Stephens, D.\ C.\ Schmidt, D.\ Seweryniak, B.\ J.\ Varley,
S.\ Vincent, K.\ Vetter, J.\ C.\ Waddington, R.\ Wadsworth,
D.\ Ward, A.\ N.\ Wilson, A.\ V.\ Afanasjev, S.\ Frauendorf,
I.\ Ragnarsson and R.\ Wyss, Proc.\ Int.\ Conf. on ``Frontiers 
of Nuclear Structure'', (Berkeley, California, 2002), AIP 
Conf. Proc. v. 656, Eds. P. Fallon and R. Clark, (Melville, New York, 
2003) p.\ 261.

%
%
\bibitem{72Kr-epj} N.\ S.\ Kelsall, C.\ E.\ Svensson, S.\ Fischer,
D.\ E.\ Appelbe, R.\ A.\ E.\ Austin, D.\ P.\ Balamuth,  G.\ C.\ Ball, J.\ A.\ Cameron,
M.\ P.\ Carpenter, R.\ M.\ Clark, M.\ Cromaz, M.\ A.\ Delaplanque, R.\ M.\ Diamond,
P.\ Fallon, D.\ F.\ Hodgson, R.\ V.\ F.\ Janssens, D.\ G.\ Jenkins, G.\ J.\ Lane,
C.\ J.\ Lister, A.\ O.\ Macchiavelli, C.\ D.\ O'Leary, D.\ G.\ Sarantites,
F.\ S.\ Stephens, D.\ C.\ Schmidt, D.\ Seweryniak, K.\ Vetter, J.\ C.\ Waddington, 
R.\ Wadsworth, D.\ Ward, A.\ N.\ Wilson, A.\ V.\ Afanasjev, S.\ Frauendorf,
I.\ Ragnarsson, Eur.\ Phys.\ Jour. A {\bf 20}, 131 (2004).

\bibitem{72Kr-Qt} G.\ de Angelis, C.\ Fahlander, A.\ Gadea, E.\ Farnea, 
W.\ Gelletly, A.\ Aprahamian, D.\ Bazzacco, F.\ Becker, P.\ G.\ Bizzeti, 
A.\ Bizzeti-Sona, F. Brandolini, D.\ de Acu\~{n}a, M.\ De Poli, J.\ Eberth, 
D.\ Foltescu, S.\ M.\ Lenzi, S.\ Lunardi, T.\ Martinez, D.\ R.\ Napoli, 
P.\ Pavan, C.\ M.\ Petrache, C.\ Rossi Alvarez, D.\ Rudolph, B.\ Rubio, 
W.\ Satu{\l}a, S.\ Skoda, P.\ Spolaore, H.\ G.\ Thomas, C.\ A.\ Ur, and R.\ Wyss,
Phys.\ Lett. B {\bf 415}, 217 (1997).


\bibitem{MMNM.82} E.\ M.\ M{\"u}ller, K.\ M{\"u}hlhans, K.\ Neerg{\aa}rd, U.\ Mosel, 
Nucl.\ Phys. {\bf A383}, 233 (1982).

\bibitem{SatW.00} S.\ Satu{\l}a and R.\ Wyss, Nucl.\ Phys. {\bf A676}, 120 (2000).

\bibitem{G.01} A.\ L.\ Goodman, Phys.\ Rev. C {\bf 63}, 044325 (2001).

\bibitem{SW.00} J.\ A.\ Sheikh and R.\ Wyss, Phys.\ Rev. C {\bf 62}, 051302(R) (2000).

\bibitem{kr72t0}
N.\ S.\ Kelsall, R.\ Wadsworth, A.\ N.\ Wilson, P.\ Fallon, A.\ O.\ Macchiavelli,
R.\ M.\ Clark, D.\ G.\ Sarantites, D.\ Seweryniak, C.\ E.\ Svensson, S.\ M.\ Vincent,
S.\ Frauendorf, J.\ A.\ Sheikh, and G.\ C.\ Ball, Phys.\ Rev. C {\bf 64}, 024309 (2001).

\bibitem{KZ.98} K.\ Kaneko and J.\ Zhang, Phys.\ Rev. C {\bf 57}, 1732 (1998).

\bibitem{FS.99} S.\ Frauendorf and J.\ A.\ Sheikh, Phys.\ Rev. C {\bf 59}, 1400 
(1999).

\bibitem{128Nd} O.\ Zeidan, D.\ J.\ Hartley, L.\ L.\ Riedinger,
W.\ Reviol, W.\ D.\ Weintraub, Y.\ Sun, Jing-ye Zhang, A.\ Galindo-Uribarri,
S.\ D.\ Paul, D.\ G.\ Sarantites, M.\ Devlin, M.\ P.\ Carpenter,
R.\ V.\ F.\ Janssens and D.\ Seweryniak, Phys.\ Rev. C {\bf 66}, 044311 (2002).

\bibitem{C-priv.02}  P.\ Chowdhury {\it et al}, to be published.

\bibitem{SWF.94} Y.\ Sun, S.\ Wen and D.\ H.\ Feng, Phys.\ Rev.\ Lett. {\bf 72}, 3483 (1994).

\end{thebibliography}
\end{document}